\documentclass{article}
\usepackage{amssymb}
\usepackage{amsfonts}
\usepackage{amsmath}
\usepackage{graphicx}
\usepackage{color}

\begin{document}

%
%
\title{New directions for gravitational wave physics via ``Millikan oil drops"*}
\author{Raymond Y. Chiao
\and Professor in the School of Natural Sciences
\and and in the School of Engineering, \\
University of California, Merced}
\maketitle

Editors' note: The following chapter has been the subject of considerable
controversy during the review process. Although one reviewer found the claim
that gravitational waves could be detected or generated by the method
proposed in this paper to be reasonable, the other reviewers found this
claim to be highly questionable. In particular, the reviewers and editors
believe that some statements in the paper may be inconsistent with the
current theory of superfluids. However, that theory could be wrong, and Dr.
Chiao's innovative work proposes an experiment based on an alternative view.
We hope that its publication here will stimulate the sort of discussion that
leads to scientific progress.
\\
\setlength{\unitlength}{1\textwidth}
\begin{picture}(0,0)
\put(0,0){\line(1,0){1}}
\end{picture}
\\
* This paper is being prepared for publication as a chapter in
\emph{Visions of Discovery: New Light on Physics, Cosmology, and Consciousness},
forthcoming from Cambridge University Press.\\

\bigskip
\bigskip
\begin{center}
\emph{And God said,
``Let there be light," and there was light. (Gen. 1:3)}
\end{center}


\section{Introduction}

In this book in honor of my beloved teacher, colleague, and friend for over
four decades, Professor Charles Hard Townes, I would like to take a fresh
look at an old problem
we had discussed on many occasions,
going back to the days when I was his graduate student at MIT. After a
visit to Joseph Weber's laboratory at the University of Maryland in the
1960s, I can still remember his critical remarks concerning the experiments
then being conducted in Weber's lab using large, massive aluminum bars. He
expressed concerns that the numbers that he calculated indicated that it
would be extremely difficult to see any observable effects, and he was
therefore worried that Weber would not be able to see any genuine signal.
Later, he expressed to me his similar worries about LIGO, especially in
light of its large scale and expense.

Here I would like to revisit the problem of \textit{generating}
gravitational radiation, which has many similarities to that of generating
electromagnetic radiation. The famous work of Gordon, Zeiger, and Townes on
the maser opened up entirely new directions in coherent electromagnetic wave
research by generating coherent microwaves by means of the quantum
mechanical principle of the stimulated emission of radiation.

Are there new ideas that might stimulate similar developments that would
open up new directions in gravitational wave research? I would like to
explore here situations in which the principle of reciprocity (i.e.,
time-reversal symmetry) demands the existence of nonnegligible quantum
back-actions of a measuring device on the gravitational radiation fields
that are being measured in a quantum mechanical context. I believe that such
quantum back-actions may allow the generation of gravitational waves.

The quantum approach taken here is in stark contrast to the classical,
test-particle approaches being taken in contemporary, large-scale
gravitational wave experiments, which are based solely on classical physics. The
back-actions of classical measuring devices such as Weber bars and large
laser interferometers on the incident gravitational fields that are being
measured are completely negligible. Hence, they can only passively detect
gravitational waves from powerful astronomical sources such as supernovae
\cite{MTW}, but they certainly cannot generate these waves.

Specifically, I would like to explore here the quantum physics of
Planck-mass-scale
``Millikan oil drops"
consisting of electron-coated superfluid helium drops at millikelvin-scale
temperatures in the presence of tesla-scale magnetic fields, as a means to
test whether or not some of the large quantum back-action effects predicted here
exist.

Recently, our ideas have shifted from the use of superfluid helium drops to
the more practical use of magnetically levitated, electrically charged
superconducting spheres, whose scattering cross section for an incoming
gravitational wave is predicted to be enormously enhanced over that for
normal, classical matter by 42 orders of magnitude \cite{Prague paper-1,Prague paper-2}.

This enormous enhancement factor arises from the ratio of the electrostatic
force to the gravitational force between two electrons and is a necessary
consequence of the uncertainty principle. When this basic quantum principle
is applied to the motion of Cooper pairs in a superconductor in the presence
of a gravitational wave, supercurrents will result because the uncertainty principle
trumps the equivalence principle whenever decoherence
is prevented from occurring due to the Bardeen, Cooper, and Schrieffer (BCS) energy
gap.\footnote{Zurek's
decoherence \cite{Zurek}
is a necessary, but not sufficient, condition for the separation of charges in
the ``Heisenberg-Coulomb" effect described
in \cite{Prague paper-1,Prague paper-2}, in which the ions and normal
electrons inside the metal undergo geodesic motion, but the Cooper pairs,
which are in the zero-momentum eigenstate of the BCS ground state, do not.}
These currents lead to an
enormous back-action on the wave, which arises from the Coulomb force
between the ensuing separated charges that strongly oppose the gravitational
tidal force of the incoming wave. Quantum back-actions are thus predicted to
lead in this case to a mirror-like reflection of the wave.

I am in the process of performing some of these experiments with my
colleagues at the new tenth campus of the University of California at Merced
in order to test some of these ideas. These quantum experiments have become
practical to perform because of important advances in ultra--low-temperature
dilution refrigerator technology. I will describe some of these experiments
below.


\section{Forces of gravity and of electricity between two electrons}

Let us first consider, using only classical, Newtonian concepts (which are
valid in the correspondence principle limit and at large distances
asymptotically, as seen by a distant observer), the forces experienced by
two electrons separated by a distance $r$ in the vacuum. Both the
gravitational force and the electrical force obey long-range, inverse-square laws.
Newton's law of gravitation states that%
\begin{equation}
\left\vert F_{G}\right\vert =\frac{Gm_{e}^{2}}{r^{2}}
\label{Newton's-inverse-square-law}
\end{equation}%
where $G$ is Newton's constant and $m_{e}$ is the mass of the electron, and
Coulomb's law states that%
\begin{equation}
\left\vert F_{e}\right\vert =\frac{e^{2}}{4\pi \varepsilon _{0}r^{2}}\text{ }
\label{Coulomb's-law}
\end{equation}%
where $e$ is the charge of the electron and $\varepsilon _{0}$ is the
permittivity of free space (I shall use SI units throughout this paper except in
Appendix B).
The electrical force between two electrons is
repulsive, but the gravitational force is attractive.

Taking the ratio of these two forces, one obtains the dimensionless ratio of
fundamental coupling constants%
\begin{equation}
\frac{\left\vert F_{G}\right\vert }{\left\vert F_{e}\right\vert }=\frac{4\pi
\varepsilon _{0}Gm_{e}^{2}}{e^{2}}\approx 2.4\times 10^{-43}
\label{Gm^2/e^2}
\end{equation}%
The gravitational force is extremely small compared to the electrical force
and is therefore usually ignored in all treatments of quantum physics.
However, it turns out that this force cannot be ignored in the case of a
superconductor interacting with an incident gravitational
wave \cite{Prague paper-1,Prague paper-2}.


\section{Gravitational and electromagnetic radiation powers emitted by two
electrons}

The above ratio of the fundamental coupling constants $4\pi \varepsilon
_{0}Gm_{e}^{2}/e^{2}$ is also the ratio of the powers of gravitational (GR)
to electromagnetic (EM) radiation emitted by two electrons separated by a
distance $r$ in the vacuum, when they undergo an acceleration $a$ and are
moving with a speed $v$ relative to each other, as seen by a distant
observer.

From the equivalence principle, it follows that dipolar gravitational
radiation does not exist \cite{MTW}. Rather, the lowest order of symmetry
of radiation permitted by this principle is quadrupolar.
General relativity predicts that the power $P_{\text{GR}}^{\text{(quad)}}$
radiated by a time-varying mass quadrupole tensor $D_{ij}$ of a periodic
system is given by \cite{MTW,Landau,Weinberg}%
\begin{equation}
P_{\text{GR}}^{\text{(quad)}}=\frac{G}{45c^{5}}\left\langle \dddot{D}%
_{ij}^{2}\right\rangle =\omega ^{6}\frac{G}{45c^{5}}\left\langle
D_{ij}^{2}\right\rangle   \label{triple-dot}
\end{equation}%
where the triple dots over $\dddot{D}_{ij}$ denote the third derivative with
respect to time of the mass quadrupole moment tensor $D_{ij}$ of the system
(the Einstein summation convention over the spatial indices $(i,j)$ for the
term $\dddot{D}_{ij}^{2}$ is being used here), $\omega $ is the angular
frequency of the periodic motion of the system, and the angular brackets
denote time averaging over one period of the motion.

Applying this formula to the periodic orbital motion of two point masses
with equal mass $m$ moving with a relative instantaneous acceleration whose
magnitude is given by $\left\vert a\right\vert =\omega ^{2}\left\vert
D\right\vert $, where $\left\vert D\right\vert $ is the magnitude of the
relative displacement of these objects, and where the relative instantaneous
speed of the two masses is given by $\left\vert v\right\vert =\omega
\left\vert D\right\vert $
(where $v \ll c$), with
all these quantities being
measured by a distant observer, one finds that Equation (\ref{triple-dot})
can be rewritten as follows:%
\begin{equation}
P_{\text{GR}}^{\text{(quad)}}=\kappa \frac{2}{3}\frac{Gm^{2}}{c^{3}}a^{2}\text{
where }\kappa =\frac{2}{15}\frac{v^{2}}{c^{2}}  \label{modified-Larmor}
\end{equation}%
The frequency dependence of the radiated power predicted by Equation (\ref%
{modified-Larmor}) scales as $v^{2}a^{2}\sim \omega ^{6}$, in agreement with
triple dot term $\dddot{D}_{ij}^{2}$ in Equation (\ref{triple-dot}). It
should be stressed that the values of the quantities $a$ and $v$ are those
being measured by an observer at infinity. The validity of Equations (\ref%
{triple-dot}) and (\ref{modified-Larmor}) has been verified by observations
of the orbital decay of the binary pulsar PSR 1913+16 \cite{Taylor1994}.

Now consider the radiation emitted by two electrons undergoing an
acceleration $a$ relative to each other with a relative speed $v$, as
observed by an observer at infinity. For example, these two electrons could
be attached to the two ends of a massless, rigid rod rotating around the
center of mass of the system like a dumbbell. The power in gravitational
radiation that they will emit is given by%
\begin{equation}
P_{\text{GR}}^{\text{(quad)}}=\kappa \frac{2}{3}\frac{Gm_{e}^{2}}{c^{3}}a^{2} \label{GR-Larmor}
\end{equation}%
where the factor $\kappa $ is given above in Equation (\ref{modified-Larmor}%
). Due to their bilateral symmetry, these two identical electrons will also
radiate quadrupolar, but not dipolar, electromagnetic radiation with a power
given by%
\begin{equation}
P_{\text{EM}}^{\text{(quad)}}=\kappa \frac{2}{3}\frac{e^{2}}{4\pi \varepsilon
_{0}c^{3}}a^{2} \label{EM-Larmor-quad}
\end{equation}%
with the same factor of $\kappa $. The reason that this is true is that any
given electron carries with it mass as well as charge as it moves, since its
charge and mass must co-move rigidly together. Therefore, two electrons
undergoing an acceleration $a$ relative to each other with a relative speed $%
v$ will emit simultaneously both electromagnetic and gravitational
radiation, and the quadrupolar electromagnetic radiation that it emits will
be completely homologous to the quadrupolar gravitational radiation that it
also emits.

It follows that the ratio of gravitational to electromagnetic radiation
powers emitted by the two-electron system is given by the same ratio of
fundamental coupling constants as that for the force of gravity relative to
the force of electricity, viz.,%
\begin{equation}
\frac{P_{\text{GR}}^{\text{(quad)}}}{P_{\text{EM}}^{\text{(quad)}}}=\frac{4\pi
\varepsilon _{0}Gm_{e}^{2}}{e^{2}}\approx 2.4\times 10^{-43}
\label{Ratio-of-powers}
\end{equation}%
Thus, it would seem at first sight hopeless to try to use any electron
system as a practical means for coupling between electromagnetic and
gravitational radiation.

Nevertheless, it should be emphasized here that although this dimensionless
ratio of fundamental coupling constants is extremely small, the
gravitational radiation emitted from the two-electron system must \emph{in
principle} exist, or else there must be something fundamentally wrong with
the experimentally well-tested inverse-square laws given by Equations (\ref%
{Newton's-inverse-square-law}) and (\ref{Coulomb's-law}).


\begin{figure}[ptb]
\label{fig01-2-oil-drops}
\par
\begin{center}
\includegraphics[width=4in]{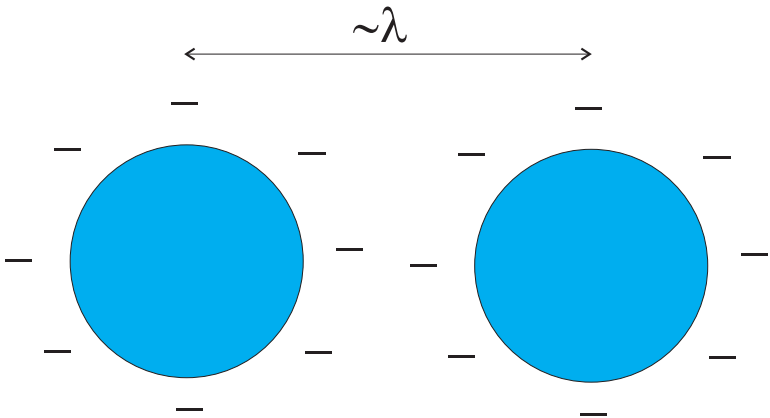}
\end{center} 
\caption{Planck-mass-scale superfluid helium drops coated with electrons on their outside surfaces and separated by approximately a microwave wavelength $\protect\lambda $, which are levitated in the presence of a
strong magnetic field.}
\end{figure}
\section{The Planck mass scale}

However, the ratio of the forces of gravity and electricity of two
``Millikan oil drops"
(see Fig.~\ref{fig01-2-oil-drops}) need
not be so hopelessly small \cite{Lamb-medal}.
%
%
%

For the purposes of an order-of-magnitude estimate, suppose that each
``Millikan oil drop"
has a single electron
attached firmly to it and contains a Planck-mass amount of superfluid
helium, viz.,%
\begin{equation}
m_{\text{Planck}}=\sqrt{\frac{\hbar c}{G}}\approx 22~\mu\text{g}
\label{Planck-mass}
\end{equation}%
where $\hbar $ is Planck's constant/2$\pi $, $c$ is the speed of light, and $%
G$ is Newton's constant. Planck's mass sets the characteristic scale at
which quantum mechanics ($\hbar $) impacts relativistic gravity ($c$, $G$).
Note that the extreme smallness of $\hbar $ compensates for the extreme
largeness of $c$ and for the extreme smallness of $G$, so that the
order of magnitude of this mass scale is \textit{mesoscopic}, and not
astronomical, in size. This suggests that it may be possible to perform some
novel \textit{nonastronomical}, tabletop experiments at the interface of
quantum mechanics and general relativity, which are accessible in any
laboratory. Such experiments will be considered here.

The forces of gravity and electricity between the two
``Millikan oil drops"
are exerted on the centers of mass
and the centers of charge of the drops, respectively. Both of these centers
coincide with the geometrical centers of the spherical drops, assuming that
the charge of the electrons on the drops is uniformly distributed around the
outside surface of the drops in a spherically symmetric manner (like in an $S
$ state). Therefore, the ratio of the forces of gravity and electricity
between the two
``Millikan oil drops"
becomes%
\begin{equation}
\frac{\left\vert F_{G}\right\vert }{\left\vert F_{e}\right\vert }=\frac{4\pi
\varepsilon _{0}Gm_{\text{Planck}}^{2}}{e^{2}}=\frac{4\pi \varepsilon
_{0}G\left( \hbar c/G\right) }{e^{2}}=\frac{4 \pi \epsilon_{0}\hbar c}{e^{2}}\approx 137 \label{137}
\end{equation}%
Now the force of gravity is approximately 137 times \textit{stronger} than
the force of electricity, so that instead of a mutual repulsion between
these two charged, massive objects, there is now a mutual attraction between
them. The sign change from mutual repulsion to mutual attraction between
these two
``Millikan oil drops"
occurs at
a critical mass $m_{\text{crit}}$ given by%
\begin{equation}
m_{\text{crit}}=\sqrt{\frac{e^{2}}{4\pi \varepsilon _{0}\hbar c}}m_{\text{%
Planck}}\approx 1.9~\mu\text{g}  \label{m_[crit]}
\end{equation}%
whereupon $\left\vert F_{G}\right\vert $ $=\left\vert F_{e}\right\vert $
and the forces of gravity and electricity balance each other in equilibrium.
The radius of a drop with this critical mass of superfluid helium, which has
a density of $\rho =0.145$ g/cm$^{3}$, is%
\begin{equation}
R=\left( \frac{3m_{\text{crit}}}{4\pi \rho }\right) ^{1/3}=146~\mu\text{m}
\end{equation}%
This is a strong hint that \textit{mesoscopic}-scale quantum effects can
lead to nonnegligible couplings between gravity and electromagnetism that
can be observed in the laboratory.

Now let us scale up the mass so that there can still occur a comparable
amount of generation of gravitational and electromagnetic radiation power
on scattering of radiation from a larger pair of
``Millikan oil drops,"
each with a larger mass $M$ and with
a larger charge $Q$, so that%
\begin{equation}
\frac{P_{\text{GR}}^{\text{(quad)}}}{P_{\text{EM}}^{\text{(quad)}}}=\frac{4\pi \varepsilon
_{0}GM^{2}}{Q^{2}}=1 \label{Larmor-power-ratio}
\end{equation}%
We shall call this the
``criticality"
condition. At
``criticality,"
equal
amounts of quadrupolar gravitational and quadrupolar electrical radiation
powers will be scattered from the two objects. The factors of $\kappa $ in
Equations (\ref{GR-Larmor}) and (\ref{EM-Larmor-quad}) still cancel out, if
the center of mass of each object co-moves rigidly together with its center
of charge. This will happen if the objects remain rigidly in their quantum
ground state. Then the scattered power from these two larger objects in the
gravitational wave channel will remain equal to that in the electromagnetic
wave channel. Of course, it will be necessary for the scattering cross
sections of gravitational waves from these objects to be nonnegligible, in
order for the scattered power to be experimentally interesting. This turns
out to be the case not only for
``Millikan oil drops,"
but also for the more practical case of a pair of
centimeter-scale, charged superconducting
spheres \cite{Prague paper-1,Prague paper-2}, which
are also levitated in the configuration shown in Figure~\ref{fig01-2-oil-drops}.

Note that any pair of objects whose masses have been increased beyond the
critical mass $m_{\text{crit}}$ can still satisfy the
``criticality"
condition, Equation (\ref{Larmor-power-ratio}%
), provided that the number of electrons on these objects is also increased
proportionately, so that their charge-to-mass ratio remains fixed, and
provided that these objects remain in their quantum mechanical ground states
during the passage of an incident GR wave, so that their motion is rigid.
Therefore, we can replace a pair of drops of superfluid helium by a pair of
spheres of superconductors, provided that these spheres are charged so that
their charge-to-mass ratio is maintained at the
``criticality" value%
\begin{equation}
\left( \frac{Q}{M}\right) _{\text{criticality}}=\sqrt{4\pi \varepsilon _{0}G}%
=8.6\times 10^{-11}\text{ C/kg}  \label{Criticality ratio}
\end{equation}%
This
``criticality"
charge-to-mass ratio
can be easily achieved experimentally. For the superconducting spheres, it
will be necessary for them to remain in the BCS ground state during the
passage of a GR wave. This can be achieved by cooling them to ultra low
temperatures.


\section{Maxwell-like equations}

To understand the calculation of the scattering cross section of the
``Millikan oil drops"
to be given below,
let us start from a useful Maxwell-like representation of the linearized
Einstein equations of general relativity due to
Wald (see Ref. \cite{Wald} and Appendix A),
which describes weak gravitational fields coupled to nonrelativistic matter in the
asymptotically flat coordinate system of a distant inertial observer:
\begin{equation}
\mathbf{\nabla \cdot E}_{G}=-\frac{\rho _{G}}{\varepsilon _{G}}
\label{Maxwell-like-eq-1}
\end{equation}%
\begin{equation}
\mathbf{\nabla \times E}_{G}=-\frac{\partial \mathbf{B}_{G}}{\partial t}
\label{Maxwell-like-eq-2}
\end{equation}%
\begin{equation}
\mathbf{\nabla \cdot B}_{G}=0  \label{Maxwell-like-eq-3}
\end{equation}%
\begin{equation}
\mathbf{\nabla \times B}_{G}=\mu _{G}\left( -\mathbf{J}_{G}+\varepsilon _{G}%
\frac{\partial \mathbf{E}_{G}}{\partial t}\right)   \label{Maxwell-like-eq-4}
\end{equation}%
where the gravitational analog of the electric permittivity of free space $%
\varepsilon _{G}$ is given by%
\begin{equation}
\varepsilon _{G}=\frac{1}{4\pi G}=1.19\times 10^{9}\text{ SI units}
\label{epsilon_G}
\end{equation}%
and where the gravitational analog of the magnetic permeability of free
space $\mu _{G}$ is given by%
\begin{equation}
\mu _{G}=\frac{4\pi G}{c^{2}}=9.31\times 10^{-27}\text{ SI units}
\label{mu_G}
\end{equation}%
Taking the curl of the gravitational analog of Faraday's law, Equation (\ref%
{Maxwell-like-eq-2}), and substituting into its right-hand side the gravitational
analog of Ampere's law, Equation (\ref{Maxwell-like-eq-4}), one obtains a
wave equation, which implies that the speed of gravitational radiation is
given by%
\begin{equation}
c=\frac{1}{\sqrt{\varepsilon _{G}\mu _{G}}}=3.00\times 10^{8}\text{ m/s}
\label{speed-of-light}
\end{equation}%
which exactly equals the vacuum speed of light. In these Maxwell-like
equations, the field $\mathbf{E}_{G}$, which is the \emph{gravitoelectric}
field, is to be identified with the local acceleration $\mathbf{g}$ of a
test particle produced by the mass density $\rho _{G}$, and the field $%
\mathbf{B}_{G}$, which is the \emph{gravitomagnetic} field produced by the
mass current density $\mathbf{J}_{G}$ and by the gravitational analog of the
Maxwell displacement current density $\varepsilon _{G}\partial \mathbf{E}%
_{G}/\partial t$, is to be identified with a time-dependent generalization
of the Lense-Thirring field of general relativity.

In addition to the speed $c$ of gravitational waves, there is another
important physical property that these waves possess, which can be formed
from the gravitomagnetic permeability of free space $\mu _{G}$ and from the
gravitoelectric permittivity $\varepsilon _{G}$ of free space, namely, the
\textit{gravitational}
characteristic impedance of free space $Z_{G}$, which is given by
\cite{Kiefer-Weber,Chiao2004-1,Chiao2004-2,Chiao2004-3}%
\begin{equation}
Z_{G}=\sqrt{\frac{\mu _{G}}{\varepsilon _{G}}}=\frac{4\pi G}{c}=2.79\times
10^{-18}\text{ SI units}  \label{Z_G}
\end{equation}%
As in electromagnetism, the characteristic impedance of free space $Z_{G}$
plays a central role in all radiation problems, such as in a comparison of
the radiation resistance of gravitational wave antennas to the value of this
impedance in order to estimate the coupling efficiency of these antennas to
free space. The numerical value of this impedance is extremely small, but
the impedance of all material objects must be much lower than this extremely
small quantity before significant power from an incident GR wave can be
appreciably scattered or reflected by these objects.

However, all classical material objects, such as Weber bars, have such a
high dissipation and such a high radiation resistance that they are
extremely poorly
``impedance matched"
to free space. They can therefore neither absorb nor scatter gravitational wave
energy efficiently \cite{Weinberg,Chiao2004-1,Chiao2004-2,Chiao2004-3}. Hence, it is a common
belief that all materials, whether classical or quantum, are essentially
completely transparent to gravitational radiation.

Macroscopically coherent quantum matter (e.g., a quantum Hall fluid) can be
an exception to this general rule, however, since it can be quantized so as
to have a strictly zero dissipation. In the quantum Hall effect, this
``quantum dissipationlessness"
arises from
the large size of the energy gap $E_{\text{gap}}=\hbar \omega _{\text{cycl}}$,
where $\omega _{\text{cycl}}$ is the electron cyclotron frequency, when $E_{%
\text{gap}}$ is compared with the small size of the thermal fluctuations due
to $k_{\text{B}}T$ at very low temperatures. The energy gap $E_{\text{gap}}$ is
like the BCS gap of superconductors \cite{Tinkham}. As in superconductors,
because of the absence of excitations with energies within the energy gap, the
scattering of the electrons in the quantum Hall fluid by phonons,
impurities, etc., in the material is exponentially suppressed, and the
quantum many-body system thus becomes dissipationless. For example,
persistent currents in annular rings of superconductors have been observed
to have lifetimes longer than the age of the universe.

Instead of discussing superconductors here, however, I focus instead
on quantum Hall fluids. (For the more practical case of superconductors, see
our work in Refs. \cite{Prague paper-1,Prague paper-2}.)


\section{Specular reflection of gravitational waves by a quantum Hall fluid}

A quantum Hall fluid consists of a two-dimensional electron gas that forms
at very low temperatures in the presence of a very strong magnetic field. In
solid-state physics, a quantum Hall fluid forms due to the electrons trapped
at the interface between two semiconductors, such as gallium arsenide and
gallium-aluminum arsenide, when the sample is cooled down to
millikelvin-scale temperatures in the presence of tesla-scale magnetic
fields. Experimental evidence that the quantum Hall fluid is dissipationless
comes from the fact that their quantum Hall plateaus are extremely flat. For
example, in the
``integer"
effect, the
transverse Hall resistance is quantized in exact integer multiples of $%
h/e^{2}$, but the longitudinal Hall resistance, which is responsible for
dissipation, is quantized to become exactly zero \cite{Prange}.

However, I  consider here the quantum Hall fluid that forms on
the surface of a superfluid helium drop. Impurity, phonon, roton, ripplon,
etc., scattering of the electrons moving on the surface of the drop is
exponentially suppressed because of the essentially perfect superfluidity of
liquid helium at millikelvin-scale temperatures. Thus, the electrons can
slide frictionlessly along the surface of a
``Millikan oil drop."
Since the electrons reside in a thin layer at a
very small distance of approximately 80 \AA\ away from the surface, which is
much smaller than the typical centimeter-scale size of the drops to be used
in the proposed experiments, locally the electronic motion is planar and can
be well approximated by the two-dimensional motion of an electron gas on a
frictionless dielectric plane (see
Appendix B).

One important consequence of the zero-resistance property of a quantum Hall
fluid is that a mirror-like reflection of electromagnetic waves can occur at
a planar interface between the vacuum and the fluid. This reflection is
similar to that which occurs when an incident electromagnetic wave
propagates down a transmission line with a characteristic impedance $Z$,
which is then terminated by means of a resistor $R$ whose value is close to
zero. The reflection coefficient ${\mathcal{R}}$ of the wave from such a
termination is given by%
\begin{equation}
{\mathcal{R}}=\left\vert \frac{Z-R}{Z+R}\right\vert ^{2}\rightarrow 100\%%
\text{ when }R\rightarrow 0
\label{Reflection-from-transmission-line}
\end{equation}%
which approaches arbitrarily close to 100\% when the resistance vanishes.
When the resistance $R=0$, low-frequency electromagnetic radiation fields
are
``shorted out"
by the resistor $R$,
and specular reflection occurs.

From the Maxwell-like Equations (\ref{Maxwell-like-eq-1}) through (\ref%
{Maxwell-like-eq-4}) and the boundary conditions that follow from
them,\footnote{Recall
\label{Hossenfelder-fn}
the boundary conditions that follow from
Maxwell's equations for electromagnetism. Consider for simplicity a planar
boundary. The local normal component of the magnetic field must be
continuous across the boundary (this comes from the Maxwell equation $%
\mathbf{\nabla \cdot B}=0$ applied to a small pillbox that straddles the
boundary), and the local tangential component of the magnetic field must
have a discontinuous jump across the boundary due to surface currents
flowing at the boundary (this comes from the Maxwell equation $\mathbf{%
\nabla \times B}=\mu _{0}\mathbf{J}$, where $\mathbf{J}$ is the electric
current density, applied to a small rectangular loop that straddles the
boundary). For a quantum Hall fluid moving frictionlessly on the surface of
superfluid helium, the surface resistance of the electrons on the surface is
strictly zero. This, in conjunction with the Lorentz force law, leads to
specular reflection of EM waves from the boundary for one circular
polarization, as
is shown in Appendix B.
But each electron carries mass as
well as charge with it when it moves. Therefore, a strictly zero surface
resistance in the electrical sector implies a strictly zero surface
resistance in the gravitational sector. The gravitational Maxwell-like
equations lead to the same local normal and tangential boundary conditions
for the gravitomagnetic field in the gravitational sector as the ones for
the electromagnetic sector. Thus, specular reflection of GR waves at
microwave frequencies should also occur below the cyclotron frequency.
While it is
true that most of the mass is in the interior of a
``Millikan oil drop,"
for the validity of the specular
boundary conditions, it is the linear response of the electrons on the
\textit{surface} of the drop to the gravitational radiation fields that is
crucial.}
it follows that an analogous reflection
of a gravitational plane wave from a planar interface of the vacuum with the
quantum Hall fluid should exist, whose reflection coefficient ${\mathcal{R}}_{G}$ is
given by%
\begin{equation}
{\mathcal{R}}_{G}=\left\vert \frac{Z_{G}-R_{G}}{Z_{G}+R_{G}}\right\vert
^{2}\rightarrow 100\%\text{ when }R_{G}\rightarrow 0
\label{Reflection-from-vacuum-superconductor-interface}
\end{equation}%
This counterintuitive result arises from the fact that the quantum Hall
fluid can, under certain circumstances, possess a strictly zero dissipation,
and therefore an equivalent mass-current resistance $R_{G}$ that can also be
strictly zero, as compared to the characteristic impedance of free space $%
Z_{G}$ $=2.79\times 10^{-18}$ SI units given by Equation (\ref{Z_G}).
Although the gravitational impedance of free space $Z_{G\text{ }}$ is an
extremely small quantity, it is still a finite quantity. However, the
dissipative resistance of a quantum Hall fluid is quantized and can
therefore be \textit{exactly} zero. When the resistance $R_{G}=0$,
low-frequency incident gravitational radiation fields are
``shorted out"
by $R_{G}$, and specular reflection occurs.

It may be objected that in Equation (\ref%
{Reflection-from-vacuum-superconductor-interface}) it is unclear exactly
how the thickness of the quantum Hall fluid compares in size relative to any
relevant
``penetration-depth"
length
scales, and also that this equation fails to take into account the
frequency-dependent complex impedance of the quantum Hall fluid. When
properly taken into account, it could have turned out that these effects would
have made the reflectivity ${\mathcal{R}}_{G}$ negligibly small. However,
when they are properly taken into account
(see Appendix C),
the result is
that although the reflectivity ${\mathcal{R}}_{G}$ is not strictly unity,
it can nevertheless be nonnegligible. The reflectivity ${\mathcal{R}}_{G}$
for gravitational waves need only be of the order of unity, and not
strictly unity, to be experimentally interesting.

Hence, it follows that under certain circumstances to be spelled out below,
specular reflection of gravitational waves can occur from a quantum Hall
fluid, just as from superconductors \cite{Prague paper-1,Prague paper-2}.
Therefore, mirrors
for gravitational radiation in principle can exist. Curved mirrors can focus
this radiation, and Newtonian telescopes for gravitational waves can
therefore in principle be constructed. In the case of scattering of
gravitational waves from the
``Millikan oil drops,"
the above specular reflection condition implies
hard-wall boundary conditions at the surfaces of these spheres, so that the
scattering cross section of these waves from a pair of large spheres can be
geometric (i.e., hard sphere) in size.

However, one cannot tell whether these statements about specular reflection
of gravitational radiation from quantum Hall fluids are true
experimentally without the existence of a source and a detector for such
radiation. The quantum transducers based on
``Millikan oil drops"
to be discussed in more detail below may provide
the needed source and detector.

Although we have been focusing in the above discussion on the case of the
quantum Hall fluid that forms on
``Millikan oil drops,"
we should remark that specular reflection of
gravitational waves should also occur from a vacuum-superconductor
interface. In addition to our recent theoretical
work \cite{Prague paper-1,Prague paper-2},
this conclusion may possibly follow from the recent potentially very
important experimental discovery \cite{Tajmar-1,Tajmar-2,Tajmar-3} (which of course needs
independent confirmation) that in an angularly accelerating superconductor,
such as a niobium ring rotating with a steadily increasing angular velocity,
there seems to be an enormous enhancement of the gravitomagnetic field $%
\mathbf{B}_{G}$. As a result of the angular acceleration of the niobium
ring, a steadily increasing gravitational analog of the
London moment in the form of a very large $\mathbf{B}_{G}$ field inside the
ring seems to arise, which is increasing linearly in time. The gravitational analog of
Faraday's law, Equation (\ref{Maxwell-like-eq-2}), then implies the
generation of loops of the gravitoelectric field $\mathbf{E}_{G}$ inside
the hole of the ring, which can be detected by sensitive accelerometers. The
gravitomagnetic field $\mathbf{B}_{G}$ is thus inferred to be many orders
of magnitude greater than what one would expect classically as a result of the mass
current associated with the rigid rotation of the ionic lattice of the ring.
These observations may have recently been confirmed by replacing the
electromechanical accelerometers with laser gyros \cite{Tajmar2}.

A tentative theoretical interpretation of these recent experiments is that
the coupling constant $\mu _{G}$, which couples the mass currents of the
superconductor to the gravitomagnetic field $\mathbf{B}_{G}$, is somehow
greatly enhanced as a result of the presence of the macroscopically coherent quantum
matter in niobium. This enhancement can be understood phenomenologically in
terms of a ferromagnetic-like enhancement factor $\kappa _{G}^{\text{(magn)}%
} $, which enhances the gravitomagnetic coupling constant \emph{inside the
medium} as follows:%
\begin{equation}
\mu _{G}^{\prime }=\kappa _{G}^{\text{(magn)}}\mu _{G}  \label{mu-G-prime}
\end{equation}%
where $\kappa _{G}^{\text{(magn)}}$ is a positive number much larger than
unity. This ferromagnetic-like enhancement factor $\kappa _{G}^{\text{(magn)}%
}$ is the gravitational analog of the magnetic permeability constant $\kappa
_{m}$ of ferromagnetic materials in the standard theory of electromagnetism.

The basic assumption of this phenomenological theory is that of a \emph{%
linear response} of the material medium to weak applied gravitomagnetic
fields;\footnote{The response of the medium must be not only \emph{%
linear} in the amplitude of the weak applied gravitational radiation fields,
but also \emph{causal}. Hence, the real and imaginary parts of
the linear response function $\kappa _{G}^{\text{(magn)}}(\omega )$, as a
function of frequency $\omega $ of the gravitational wave, must obey
Kramers-Kronig relations similar to those given by Equations (4) and (5) of
Ref. \cite{Chiao2004-1}.}
that is to say, whatever the fundamental
explanation is of the large observed positive values of $\kappa _{G}^{\text{%
(magn)}}$, the medium produces an enhanced gravitomagnetic field $\mathbf{B}%
_{G}$ that is \emph{directly proportional} to the mass current density $%
\mathbf{J}_{G}$ of the ionic lattice. For weak fields, this is a reasonable
assumption. However, it should be noted that this phenomenological
explanation based on Equation (\ref{mu-G-prime}) is different from the
theoretical explanation based on Proca-like equations for gravitational
fields with a finite graviton rest mass, which was proposed by the
discoverers of the effect in Refs. \cite{Tajmar-1,Tajmar-2,Tajmar-3}.

Nevertheless, it is natural to consider introducing the phenomenological
Equation (\ref{mu-G-prime}) to explain the observations, since a large
enhancement factor $\kappa _{G}^{\text{(magn)}}$ due to the material medium
is very similar to its analog in magnetism, which explains, for example, the
large ferromagnetic enhancement of the inductance of a solenoid by a
magnetically soft, permeable iron core with permeability
$\kappa _{m} \gg 1$
that arises from the alignment of electron spins inside the iron. This
spin-alignment effect leads to the large observed values of the magnetic
susceptibility of iron, like those utilized in mu metal shields. Just as in
the case of the iron core inserted inside a solenoid, where the large
enhancement of the solenoid's inductance disappears above the Curie
temperature of iron, it was observed in these recent experiments that the
large gravitomagnetic enhancement effect disappears above the
superconducting transition temperature of niobium.

If the tentative phenomenological interpretation given by Equation (\ref%
{mu-G-prime}) of these experiments turns out to be correct, one important
consequence of the large resulting values of $\kappa _{G}^{\text{(magn)}}$
is that a mirror-like reflection should occur at a planar
vacuum-superconductor interface, where the refractive index of the
superconductor has an abrupt jump from unity to a value given by%
\begin{equation}
n_{G}=\left( \kappa _{G}^{\text{(magn)}}\right) ^{1/2}
\label{refractive-index}
\end{equation}

However, it should be immediately emphasized here that only positive masses
are observed to exist in nature, and not negative ones. Hence, gravitational
analogs of permanent electric dipole moments do not exist. It follows that
the gravitational analog $\kappa _{G}^{\text{(elec)}}$ of the usual
dielectric constant $\kappa _{e}$ for all kinds of matter, whether classical
or quantum, in the Earth's gravitoelectric
field $\mathbf{E}_{G}=\mathbf{g}$ %
cannot differ from its vacuum value of unity---that is,%
\begin{equation}
\kappa _{G}^{\text{(elec)}}\equiv \varepsilon _{G}^{\prime }/\varepsilon
_{G}=1 \label{gravito-dielectric-constant=1}
\end{equation}%
exactly. Hence, one cannot screen out, even partially, the gravitoelectric
DC gravitational fields like the Earth's gravitational field using
superconducting Faraday cages, in an
``antigravity"
effect. In particular, the local value of
the acceleration $\mathbf{g}$ due to Earth's gravity is not at all affected
by the presence of nearby matter with large $\kappa _{G}^{\text{(magn)}}$.

The gravitational analog of Ampere's law combined with Wald's gravitational
analog of the Lorentz force law
(see Ref. \cite{Wald}, section 4.4)
\begin{equation}
\mathbf{F}_{G}=m\left( \mathbf{E}_{G}+4\mathbf{v\times B}_{G}\right) \label{Loretntz-Force-Law}
\end{equation}%
where $\mathbf{F}_{G}$ is the force on a test particle with mass $m$ and
velocity $\mathbf{v}$ (all quantities as seen by the distant inertial
observer), leads to the fact that a \textit{repulsive} component of force
exists between two parallel mass currents traveling in the same direction,
whereas two parallel electrical currents traveling in the same direction
\textit{attract} each other.
A repulsive \textit{gravitomagnetic%
} gravitational force follows from the negative sign in front of the mass
current density $\mathbf{J}_{G}$ in Equation (\ref{Maxwell-like-eq-4}),
which is necessitated by the conservation of mass, since upon taking the
divergence of Equation (\ref{Maxwell-like-eq-4}) and combining it with
Equation (\ref{Maxwell-like-eq-1}) (whose negative sign in front of the mass
density $\rho _{G}$ is fixed by Newton's law of gravitation, where all
masses \emph{attract} each other), one must obtain the continuity equation
for mass---that is,
\begin{equation}
\mathbf{\nabla \cdot J}_{G}+\frac{\partial \rho _{G}}{\partial t}=0
\label{continuity}
\end{equation}%
where $\mathbf{J}_{G}$ is the mass current density and $\rho _{G}$ is the
mass density. Moreover, the negative sign in front of the mass current
density $\mathbf{J}_{G}$ in the gravitational analog of Ampere's law,
Equation (\ref{Maxwell-like-eq-4}), implies an \textit{anti-Meissner}
effect, in which the lines of the $\mathbf{B}_{G}$ field, instead of being
expelled from the superconductor as in the usual Meissner effect, are
pulled tightly into the interior of the body of the superconductor, whenever
$\kappa _{G}^{\text{(magn)}}$ is a large, positive number.

However, it should again be stressed that what is being tentatively proposed
here in this phenomenological scenario does not at all imply an
``antigravity"
effect, in which the
Earth's gravitational field is somehow partially screened out by the
so-called
Podkletnov effect,
where it
was claimed that rotating superconductors reduce by a few percent the
gravitoelectric field $\mathbf{E}_{G}=\mathbf{g}$, i.e., the local
acceleration of all objects due to Earth's gravity, in their vicinity.
Experiments attempting to reproduce this effect have failed to do so
\cite{Tajmar-1,Tajmar-2,Tajmar-3}. The nonexistence of the
Podkletnov effect
would be consistent with the above
phenomenological theory, since \textit{longitudinal} gravitoelectric fields
cannot be screened under any circumstances; however, \textit{transverse}
radiative gravitational fields can be \textit{reflected} by
supercurrents
in coherent quantum matter.

Very large values of $\kappa _{G}^{\text{(magn)}}$ for superconductors would
imply that the index of refraction for gravitational plane waves in these
media would be considerably larger than unity---that is,%
\begin{equation}
n_{G}=\left( \kappa _{G}^{\text{(magn)}}\right) ^{1/2}\gtrsim 1
\label{index>>1}
\end{equation}%
The Fresnel reflection coefficient ${\mathcal{R}}_{G}$ of gravitational
waves normally incident on the vacuum-superconductor interface would
therefore become%
\begin{equation}
{\mathcal{R}}_{G}=\left\vert \frac{n_{G}-1}{n_{G}+1}\right\vert ^{2}\simeq
\text{order of unity}  \label{Fresnel-reflection}
\end{equation}%
and could thus be large enough to be experimentally interesting. Again, but
for different reasons from those given in \cite{Prague paper-1,Prague paper-2},
Equation (\ref%
{Fresnel-reflection}) would imply mirror-like reflection of these waves from
superconducting surfaces
(see Appendix C).
It should be noted that large
values of the ferromagnetic-like enhancement factor $\kappa _{G}^{\text{%
(magn)}}$, of the index of refraction $n_{G}$, and of the reflectivity ${%
\mathcal{R}}_{G}\,$ are not forbidden by the principle of equivalence,
which has been checked experimentally with extremely high accuracy, but only
within the gravitoelectric sector of gravitation.

However, although interesting and possibly very important, the above
discussion concerning superconductors as mirrors for gravitational waves is
only secondary to the primary purpose of this paper, which is to present the
case for the possibility of efficient quantum transducers via
``Millikan oil drops."
Nevertheless,
superconducting transducers based on the same principles as those of the
``Millikan oil drops,"
to be described
below, should also exist.


\section{``Millikan oil drops" described in more detail}

Let the oil of the classic Millikan oil drops be replaced with superfluid
helium ($^{4}$He) with a gravitational mass of approximately the
Planck mass scale, and let these drops be levitated in the presence of
strong, tesla-scale magnetic fields.

The helium atom is diamagnetic, and liquid helium drops have successfully
been magnetically levitated in an anti-Helmholtz magnetic trapping
configuration \cite{Weilert1996-1,Weilert1996-2}. As a result of its surface tension,
the surface of
a freely suspended, isolated, ultracold superfluid drop is ideally smooth,
i.e., atomically perfect, in the sense that there are no defects (such as
dislocations on the surface of an imperfect crystal) that can trap and
thereby localize the electron. The absence of any scattering centers for the
electrons on the surface of the superfluid helium of a
``Millikan oil drop"
implies that the electrons can move
frictionlessly, and hence dissipationlessly, over its surface.

When an electron approaches a drop, the formation of an image charge inside
the dielectric sphere of the drop causes the electron to be attracted by the
Coulomb force to its own image. As a result, it is experimentally observed
that the electron is bound to the outside surface of the drop in a
hydrogenic ground state. The binding energy of the electron to the surface
of liquid helium has been measured using millimeter-wave spectroscopy to be
8 K,\footnote{See Refs. \cite{Grimes2-1,Grimes2-2}.
\label{Grimes2-fn}
In the ground state of the system, the electron resides on the
\emph{outside} surface of a superfluid helium drop, and not within the \emph{%
inside} volume of the drop. When the electron is forced to be within the
interior of the drop, it will form a bubble with a radius of approximately
1 nm, as a result of the balancing of an outward Pauli pressure with the
surface tension of the superfluid (see Ref. \cite{Donnelly-1967}).
The bubble will then rise to the surface, driven by the Coulomb force of
attraction to its own image charge induced in the surface. It will then
burst through the surface to uniformly coat the drop with one electron
charge on its outside surface. The electron will be in an $S$-state to minimize the energy of the system. This then is the ground state of the
system.}
which is quite large compared to the
millikelvin-scale temperatures for the proposed experiments. Hence, the
electron is tightly bound to the outside surface of the drop so that the
radial component of its motion is frozen, but when the drop becomes a
superfluid, the electron is free to move frictionlessly tangentially on the
surface, and thus free to become delocalized over the entire surface.

Such a
``Millikan oil drop"
is a macroscopically phase-coherent quantum object. In its ground state, which
possesses a single, coherent quantum mechanical phase throughout the
interior of the
superfluid,\footnote{Note that the quantum mechanical ground-state
wave function (or complex order parameter) must remain \textit{single valued}
(according to a distant inertial observer) globally at all times everywhere
inside the interior of the system during the passage of a gravitational
wave. This is another aspect of the ``quantum rigidity" of a quantum fluid
in its response to the gravitational wave.}
the drop possesses a zero
circulation quantum number (i.e., contains no quantum vortices), with one
unit (or an integer multiple) of the charge quantum number. As a result of
the drop being at ultra low temperatures, all degrees of freedom other than
the center-of-mass degrees of freedom are frozen out, so that
a zero-phonon M\"{o}ssbauer-like effect results, in which the entire mass of the
drop moves rigidly as a single unit in response to radiation fields (see
below). Therefore, the center of mass of the drop will co-move with the
center of charge. In addition, since it remains adiabatically in the ground state
during perturbations as a result of these weak radiation fields, the
``Millikan oil drop"
possesses properties of
``quantum rigidity"
and
``quantum dissipationlessness,"
which are
the two most important quantum properties for achieving a high coupling
efficiency for gravitational wave antennas \cite{Chiao2004-1,Chiao2004-2,Chiao2004-3}.

Note that two spatially separated
``Millikan oil drops"
with the same mass and charge have the correct
bilateral symmetry in order to couple to quadrupolar gravitational
radiation, as well as to quadrupolar electromagnetic radiation in the TEM$_{%
\text{11}}$ mode. The coupling of the drops to the electromagnetic TEM$_{%
\text{00}}$ mode, however, vanishes as a result of symmetry. When they are separated
by a distance on the order of a wavelength, they should become an efficient
quadrupolar antenna capable of generating, as well as detecting,
gravitational radiation.

\begin{figure}[ptb]
\label{fig02-gr-to-em-wave-transducer}
\par
\begin{center}
\includegraphics[width=4in]{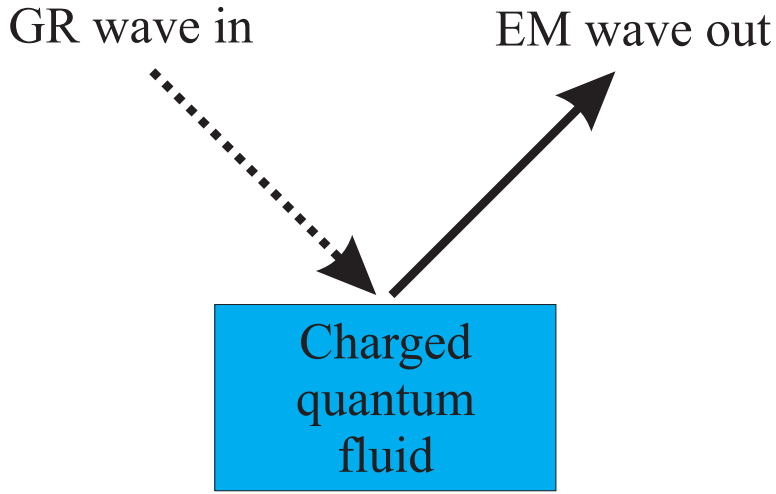}
\end{center} 
\caption{``Charged quantum fluid" is a quantum transducer consisting of a pair of ``Millikan oil drops" in a strong magnetic field, which converts a gravitational (GR) wave into an electromagnetic (EM) wave. A pair of charged superconducting spheres can also be used as
such a transducer.}
\end{figure}

%
%
%
%
%


\section{A pair of ``Millikan oil drops'' as a transducer}

Now imagine placing a pair of levitated
``Millikan oil drops"
separated by approximately a microwave wavelength
inside a black box, which represents a quantum transducer that can convert
gravitational (GR) waves into electromagnetic (EM) waves
(see Fig.~\ref{fig02-gr-to-em-wave-transducer}).
%
%
This kind of transducer action is similar to
that of the tidal force of a gravitational wave passing over a pair of
charged, freely falling objects orbiting the Earth, which can in principle
convert a GR wave into an EM wave \cite{Lamb-medal}. Such transducers are
linear, reciprocal devices.

By time-reversal
symmetry,\footnote{Time-reversal symmetry under the \emph{global} operation of
time reversal includes here the reversal of the direction of the applied DC
magnetic fields.}
the reciprocal process, in which another pair of
``Millikan oil drops"
converts an EM wave back into a GR wave, must occur with the same efficiency
as the forward process, in which a GR wave is converted into an EM wave by
the first pair of
``Millikan oil drops."
The time-reversed process is important because it allows the \emph{generation%
} of gravitational radiation and therefore can become a practical source of
such radiation. The radiation reaction or back-action by the EM fields on
the GR fields via these coherent quantum drops leads necessarily to a
nonnegligible reciprocal process of the generation of these fields. These
actions must be mutual ones between these two kinds of radiation fields.

\begin{figure}[ptb]
\label{fig03-Hertz-4b}
\par
\begin{center}
\includegraphics[width=5in]{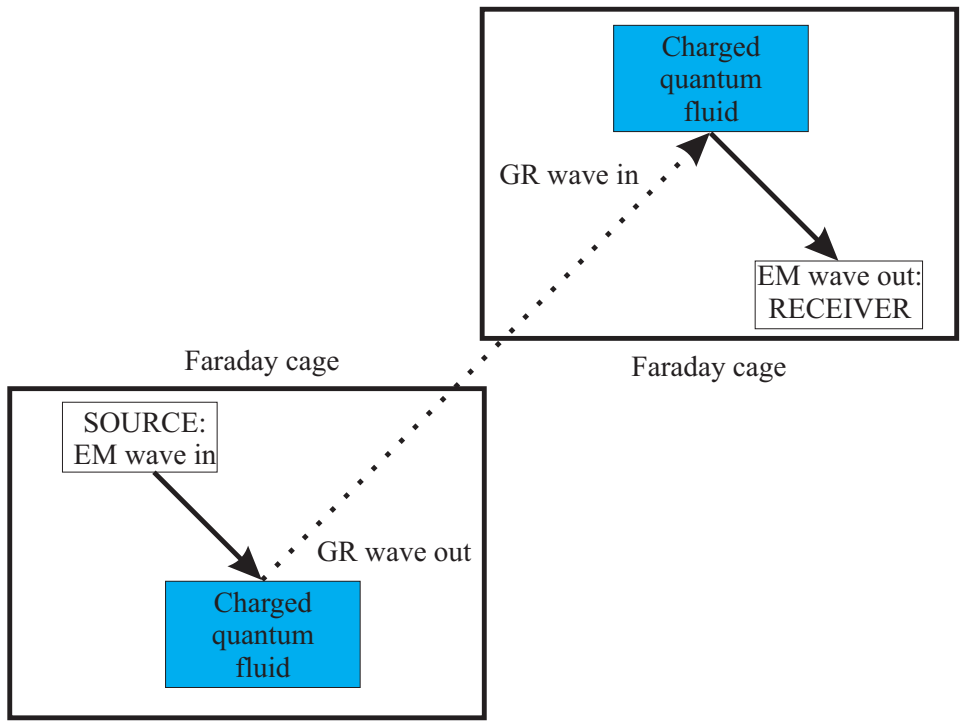}
\end{center} 
\caption
{A Hertz-like experiment, in which EM waves are converted by the lower-left quantum transducer (``Charged quantum fluid") into GR waves at the source, and the GR waves thus generated are back-converted back into EM waves by the upper-right quantum transducer at the receiver. Communication by EM waves is prevented by the normal (i.e., nonsuperconducting) Faraday cages.}
\end{figure}

This raises the possibility of performing a Hertz-like experiment, in which
the time-reversed quantum transducer process becomes the source, and its
reciprocal quantum transducer process becomes the receiver of GR waves
(see Fig.~\ref{fig03-Hertz-4b}).
%
Faraday cages consisting of nonsuperconducting metals prevent the
transmission of EM waves, so that only GR waves, which can easily pass
through all classical matter such as the normal (i.e., dissipative) metals
of which standard, room-temperature Faraday cages are composed, are
transmitted between the two halves of the apparatus that serve as the source
and the receiver, respectively. Such an experiment would be practical to
perform using standard microwave sources and receivers, provided that the
scattering cross sections and the transducer conversion efficiencies of the
two
``Millikan oil drops"
turn out not to
be too small.

%
%
%

%


\section[M{\"{o}}ssbauer-like response of ``Millikan oil drops"]{M{\"{o}}ssbauer-like
response of ``Millikan oil drops" in strong magnetic fields to radiation fields}

Let a pair of levitated
``Millikan oil drops"
be placed in strong, tesla-scale magnetic fields,
and let the drops be separated by a distance on the order of a microwave
wavelength, which is chosen so as to satisfy the impedance-matching
condition for a good quadrupolar microwave antenna.

Now let a beam of electromagnetic waves in the Hermite-Gaussian TEM$_{11}$
mode \cite{Yariv1967}, which has a quadrupolar transverse field pattern
homologous to that of a gravitational plane wave, impinge at a 45$^{\circ }$
angle with respect to the line joining these two charged objects. Such a
mode has been successfully generated using a
``T"-shape microwave antenna \cite{Chiao2004-1,Chiao2004-2,Chiao2004-3}. As a result
of being thus irradiated, the pair of
``Millikan oil drops"
will be driven into relative motion in an
antiphased manner, so that the distance between them will oscillate
sinusoidally with time, according to an observer at infinity. Thus, the
simple harmonic motion of the two drops relative to one another (as seen by
this observer) produces a time-varying mass quadrupole moment at the same
frequency as that of the driving electromagnetic wave. This oscillatory
motion will in turn scatter (in a linear scattering process) the incident
electromagnetic wave into gravitational and electromagnetic scattering
channels with comparable powers, provided that the ratio of quadrupolar
radiation powers is that given by the
``criticality"
condition, Equation (\ref{Larmor-power-ratio}%
)---that is, this ratio
is of the order of unity, which will be the case if the charge-to-mass
ratio of the drops is given by the
``criticality"
ratio, Equation (\ref{Criticality ratio}).
The reciprocal scattering process will also have a power ratio of the order
of unity.

The M{\"{o}}ssbauer-like response of
``Millikan oil drops"
will now be discussed in more detail. Imagine what
would happen if one were to replace an electron in the vacuum with a single
electron that is firmly attached to the outside surface of a drop of
superfluid helium in the presence of a strong magnetic field and at
ultra low temperatures, so that the system of the electron and the
superfluid, considered as a single quantum entity like that of a
``gigantic atom,"
would form a single, macroscopic quantum ground
state.\footnote{This single quantum entity can be viewed as if it
were a gigantic atom in which the
usual atomic nucleus is replaced by the superfluid helium drop, and the
usual electronic cloud surrounding the atomic nucleus is replaced by the
electrons on the surface surrounding the drop. The large energy
gap (Eq. (\ref{Cyclotron-gap})) arising from the large applied magnetic
field is what makes this gigantic atom extremely rigid and dissipationless
at low temperatures. A pair of such gigantic atoms forms a gigantic diatomic
molecule. If the charges and masses of the two drops are slightly different
from each other, such a gigantic diatomic molecule will form an entangled
state of charge and mass in its ground state at sufficiently low
temperatures.}
Such a quantum system
can possess a sizable gravitational mass. For the case of many electrons
attached to a large, massive drop, where a quantum Hall fluid forms on the
outside surface of the drop in the presence of a strong magnetic field,
a Laughlin-like ground state results, which is the many-body state of
an incompressible quantum fluid \cite{Laughlin}. The property of quantum
incompressibility of such a fluid is equivalent to the property of
``quantum rigidity,"
which is one
necessary requirement for achieving high efficiency in
gravitational radiation antennas, as was pointed out in
Refs.
\cite{Chiao2004-1,Chiao2004-2,Chiao2004-3}.
Like superfluids and superconductors, this fluid is also frictionless (i.e.,
dissipationless). This fulfills the condition of
``quantum dissipationlessness,"
which is another necessary
requirement for the successful construction of efficient gravitational wave
antennas \cite{Chiao2004-1,Chiao2004-2,Chiao2004-3}.

In the presence of strong, tesla-scale magnetic fields, an electron is
prevented from moving at right angles to the local magnetic field line
around which it is executing tight cyclotron orbits. The result is that the
surface of the drop, to which the electron is tightly bound, cannot undergo
low-frequency liquid-drop deformations, such as the oscillations between the
prolate and oblate spheroidal configurations of the drop that would occur
at low frequencies in the absence of the magnetic field. After the drop has
been placed into tesla-scale magnetic fields at millikelvin-scale operating
temperatures, both the single- and many-electron drop systems will be
effectively frozen into the ground state, since the characteristic energy
scale for electron cyclotron motion in tesla-scale fields is on the order of
kelvins. As a result of the tight coupling of the electron(s) to the outside surface
of the drop, also on the scale of kelvins, this would effectively freeze out
all low-frequency shape deformations of the superfluid drop.

Since all internal degrees of freedom of the drop, such as its microwave
phonon excitations, will also be frozen out at sufficiently low
temperatures, the charge and the entire mass of the
``Millikan oil drop"
will co-move rigidly together as a
single unit, in a zero-phonon, M\"{o}ssbauer-like response to applied
radiation fields with frequencies below the cyclotron frequency. This is a
result of the elimination of all internal degrees of freedom by the
Boltzmann factor at sufficiently low temperatures, so that the system stays
in its ground state, and only the external degrees of freedom of the drop,
consisting only of its center-of-mass motions, remain.

The criterion for this zero-phonon, or M\"{o}ssbauer-like, mode of response
of the electron-drop system is that the temperature of the system is
sufficiently low, so that the probability for the entire system to remain in
its ground state without even a single quantum of excitation of any of its
internal degrees of freedom being excited is very high---that is,%
\begin{equation}
\text{Prob. of zero internal excitation}\approx 1-\exp \left( -\frac{E_{%
\text{gap}}}{k_{\text{B}}T}\right) \rightarrow 1\text{ as }\frac{k_{\text{B}}T}{E_{\text{%
gap}}}\rightarrow 0  \label{Prob(no excitation)}
\end{equation}%
where $E_{\text{gap}}$ is the energy gap separating the ground state from
the lowest permissible excited states, $k_{\text{B}}$ is Boltzmann's constant, and $%
T$ is the temperature of the system. Then the quantum adiabatic theorem
ensures that the system will stay adiabatically in the ground state of this
quantum many-body system during adiabatic perturbations, such as those due
to weak, externally applied radiation fields with frequencies below the
cyclotron frequency. By momentum conservation, because there are no internal
excitations to take up the radiative momentum transfer, the center of mass
of the entire system must undergo recoil in the emission and absorption of
radiation. Thus, the mass involved in the response to radiation fields is the
entire mass of the whole system.

For the case of a single electron (or many electrons in the case of the
quantum Hall fluid) in a strong magnetic field, the typical energy gap is
given by%
\begin{equation}
E_{\text{gap}}=\hbar \omega _{\text{cycl}}=\frac{\hbar eB}{m} \gg k_{\text{B}}T  \label{Cyclotron-gap}
\end{equation}%
where $\omega _{\text{cycl}}=eB/m$ is the electron cyclotron frequency. This
is satisfied by the tesla-scale fields and millikelvin-scale temperatures
in the proposed experiments.


\section{Estimate of the scattering cross section}

Let $d\sigma _{a\rightarrow \beta }$ be the differential cross section for
the scattering of a mode $a$ of radiation of an incident gravitational wave
to a mode $\beta $ of a scattered electromagnetic wave by a pair of
``Millikan oil drops"
(Latin subscripts
denote GR waves, and Greek subscripts denote EM waves). Then, by time-reversal
symmetry\footnote{As previously noted, time-reversal
symmetry under the \emph{global} operation of
time reversal includes here the reversal of the direction of the applied DC
magnetic fields.}
\begin{equation}
d\sigma _{a\rightarrow \beta }=d\sigma _{\beta \rightarrow a}
\end{equation}%
Since electromagnetic and weak gravitational fields both formally obey
Maxwell's equations (apart from a difference in the signs of the source
density and the source current density; see Eqs. (\ref%
{Maxwell-like-eq-1})--(\ref{Maxwell-like-eq-4})), and since these fields
obey the same boundary conditions
(see Appendix C and the footnote on page \pageref{Hossenfelder-fn}),
the
solutions for the modes for the two kinds of scattered radiation fields must
also have the same mathematical form. Let $a$ and $\alpha $ be a pair of
corresponding solutions and $b$ and $\beta $ be a different pair of
corresponding solutions to Maxwell's equations for GR and EM modes,
respectively. For example, $a$ and $\alpha $ could represent incoming plane
waves that copropagate in the same direction, and $b$ and $\beta $
could represent scattered, outgoing plane waves that copropagate together in a different
direction. Then for a pair of drops with the
``criticality"
charge-to-mass ratio given by Equation (\ref%
{Criticality ratio}), there is an equal conversion into the two types of
scattered radiation fields in accordance with Equation (\ref%
{Larmor-power-ratio}), and therefore at
``criticality"
\begin{equation}
d\sigma _{a\rightarrow b}=d\sigma _{a\rightarrow \beta }
\end{equation}%
where $b$ and $\beta $ are corresponding modes of the two kinds of scattered
radiations.

By the same line of reasoning, for this pair of drops%
\begin{equation}
d\sigma _{b\rightarrow a}=d\sigma _{\beta \rightarrow a}=d\sigma _{\beta
\rightarrow \alpha }
\end{equation}%
It therefore follows from the principle of reciprocity (i.e., detailed
balance or time-reversal symmetry) that%
\begin{equation}
d\sigma _{a\rightarrow b}=d\sigma _{\alpha \rightarrow \beta }
\end{equation}

To estimate the size of the total cross section, it is easier to
consider first the case of electromagnetic scattering, such as the
scattering of microwaves from a pair of large drops with radii $R$ and a
separation $r$ on the order of a microwave wavelength (but with $r>2R$). The
diameter $2R$ of the drops can be made to be comparable to their separation $%
r\simeq \lambda $ (e.g., with $2\pi R=\lambda $ for the first Mie
resonance), provided that many electrons are added on their surfaces, so
that the
``criticality"
charge-to-mass
ratio is maintained (this requires the addition of 20,000 electrons for
the first Mie resonance at $\lambda =2.5$ cm, where $R=4$ mm).

For an incident EM wave of a particular circular polarization, even just a
single, delocalized electron in the presence of a strong magnetic field is
enough to produce specular reflection of this wave (see
Appendix B).
Therefore, for circularly polarized light, the two drops behave like
perfectly conducting, shiny, mirror-like spheres, which scatter light in a
manner similar to that of perfectly elastic hard-sphere scattering in
idealized billiards. The total cross section for the scattering of
electromagnetic radiation from this pair of large drops is therefore given
approximately by the geometric cross-sectional areas of two hard spheres%
\begin{equation}
\sigma _{\alpha \rightarrow \text{all }\beta }=\int d\sigma _{\alpha
\rightarrow \beta }\simeq \text{order of }\pi R^{2}
\label{Geometric-X-section}
\end{equation}%
where $R$ is the hard-sphere radius of a drop. This hard-sphere
cross section is much larger than the Thomson cross section for the
classical, \emph{localized} single free-electron scattering of
electromagnetic radiation.

However, if, as one might expect on the basis of the prevailing (but
possibly incorrect) opinion that all gravitational interactions with matter,
including the scattering of gravitational waves from all types of matter, are
completely independent of whether this matter is classical or
quantum mechanical in nature on any scale of size, and that therefore the
scattering cross section for the drops would be extremely small as it is for
the classical Weber bar, then by reciprocity the total cross section for
the scattering of electromagnetic waves from the two-drop system must also
be extremely small. In other words, if
``Millikan oil drops"
were to be essentially invisible to gravitational
radiation as is commonly believed, then by reciprocity they must also be
essentially invisible to electromagnetic radiation. To the contrary, if it
should turn out that the quantum Hall fluid on the surface of these drops
should make them behave like superconducting spheres, then the earlier
discussion in connection with Equation (\ref%
{Reflection-from-vacuum-superconductor-interface}) would imply that the
total cross section of these drops will be like that of hard-sphere
scattering, so that they certainly would not be invisible.


\section{A proposed preliminary experiment}

%

\begin{figure}[ptb]
\label{fig04-Rescaled-scattering-experiment-4}
\par
\begin{center}
\includegraphics[width=5in]{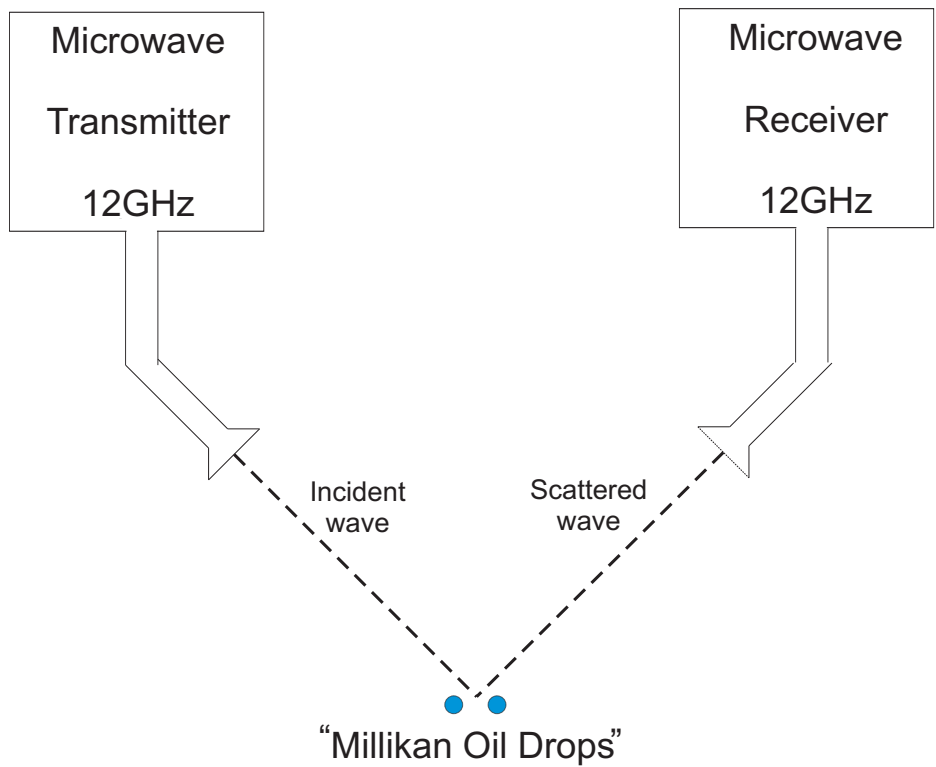}
\end{center} 
\caption{Schematic of apparatus (not to scale) to measure the scattering cross section of quadrupolar microwaves from a pair of ``Millikan oil drops" in a strong magnetic field at low temperatures.}
\end{figure}

To check the above hard-sphere scattering cross section result, we
propose first to perform in a preliminary experiment a measurement of the
purely EM scattering cross section for quadrupolar microwave radiation off
of a pair of large
``Millikan oil drops"
%
(see Fig.~\ref{fig04-Rescaled-scattering-experiment-4}).
%
An oscillator at 12 GHz emits microwaves that are prepared
in a quadrupolar TEM$_{11}$ mode and directed in a beam toward these drops,
which are placed in a large magnetic field and cooled to ultra low
temperatures.
The intensity of the scattered microwave beam generated by
the pair of drops is then measured by means of a 12 GHz heterodyne receiver,
which receives a quadrupolar TEM$_{11}$ mode. The purpose of this experiment
is to check whether the scattering cross section is indeed as large as the
geometric cross section predicted by Equations (\ref%
{Reflection-from-vacuum-superconductor-interface}), (\ref%
{Geometric-X-section}), and (\ref{Fresnel-Reflection-for-plasma}). As one
increases the temperature, one should observe the disappearance of this
enhanced scattering cross section above the quantum Hall transition
temperature or the superfluid lambda point, whichever comes first.

%
%
%

%


\section{A common misconception corrected}

In connection with the idea that an EM wave incident on a pair of drops
could generate a GR wave, a common misconception arises that the drops
are so heavy that their large inertia will prevent them from moving with any
appreciable amplitude in response to the driving EM wave amplitude. How can
they then possibly generate copious amounts of GR waves? This objection
overlooks the major role played by the principle of equivalence in the
motion of the drops, as will be explained below.

According to the equivalence principle, two tiny inertial observers, who are
undergoing free fall (i.e., who are freely floating near their respective
centers of the two
``Millikan oil drops")
would see no acceleration at all of the nearby surrounding matter of their
drop (nor would they feel any forces) as a result of the gravitational fields
arising from a gravitational wave passing over the two drops. However, when
they measure the distance separating the two drops, by means of laser
interferometry, for example, they would conclude that the other drop is
undergoing acceleration relative to their drop, as a result of the fact that the
\textit{space} between the drops is being
periodically
stretched and squeezed
by the incident gravitational wave. They would therefore further conclude
that the charges attached to the surfaces of their locally freely falling
drops would radiate electromagnetic radiation, in agreement with the
observations of the observer at infinity, who sees two charges undergoing
time-varying relative acceleration in response to the passage of the
gravitational wave.

According to the reciprocity principle, this scattering process can be
reversed in time. Under time reversal, the scattered electromagnetic wave
now becomes a wave that is incident on the drops. Again, the two tiny
inertial observers near the center of the drops would see no acceleration at
all of the surrounding matter (nor would they feel any forces) because of the
electric and magnetic fields of the incident electromagnetic wave. Rather,
they would conclude from measurements of the distance separating the two
drops that it is again the \textit{space} between the drops that is being
periodically
squeezed and stretched by the incident electromagnetic wave.
They would again further conclude that the masses associated with their
locally freely falling drops would radiate gravitational radiation, in
agreement with the observations of the observer at infinity, who sees two
masses undergoing time-varying relative acceleration in response to the
passage of the electromagnetic wave.

From this general relativistic viewpoint, which is based on the
equivalence principle, the fact that the drops might possess very large
inertias is irrelevant, since in fact the drops are not moving at all with
respect to the local inertial observer located at the center of drop.
Instead of causing motion of the drops \emph{through} space, the
gravitational fields of the incident gravitational wave are acting directly
\emph{on} space itself by
periodically
stretching and squeezing the space
in between the drops. Likewise, in the reciprocal process the very large
inertias of the drops are again irrelevant, since the electromagnetic wave
is not producing any motion at all of these drops with respect to the same
inertial observer
(see Appendix D).
Instead of causing motion of the drops
\emph{through} space, the electric and magnetic fields of the incident
electromagnetic wave are again acting directly \emph{on} space itself by
periodically
squeezing and stretching the space in between the drops. The
time-varying, accelerated motion of the drops as seen by the distant
observer that causes quadrupolar radiation to be emitted in both cases is
due to the time-varying \textit{curvature} of spacetime induced both by the
incident gravitational wave and by the incident electromagnetic wave. It
should be remembered that the space inside which the drops reside is
therefore no longer flat, so that the Newtonian concept of a
radiation-driven, local accelerated motion of a heavy drop with a large
inertia \textit{through} a fixed and flat Euclidean space is therefore no
longer valid.


\section[The strain of space produced by the drops]{The strain of space
produced by the drops for a milliwatt of GR wave power}

Another common objection to these ideas is that the strain of space produced
by a milliwatt of an electromagnetic wave is much too small to detect.
However, in the Hertz-like experiment, one is not trying to detect directly
the \textit{strain} of space (as in LIGO), but rather the \textit{power}
that is being transferred by the gravitational radiation fields from the
source to the receiver.

Let us put in some numbers. Suppose that one succeeded in completely
converting a milliwatt of EM wave power into a milliwatt of GR wave power at
the source. How big a strain amplitude of space would be produced by the
resulting GR wave? The gravitational analog of the time-averaged Poynting
vector is given by \cite{Saulson}%
\begin{equation}
\left\langle S\right\rangle =\frac{\omega ^{2}c^{3}}{32\pi G}h_{+}^{2}
\end{equation}%
where $h_{+}$ is the dimensionless strain amplitude of space for one
polarization of a monochromatic plane wave. For a milliwatt of power in such
a plane wave at 30 GHz focused by means of a Newtonian telescope to a 1 cm$%
^{2}$ Gaussian beam waist, one obtains a dimensionless strain amplitude of%
\begin{equation}
h_{+}\simeq 0.8\times 10^{-28}
\end{equation}%
This strain is indeed exceedingly difficult to detect directly. However,
it is not necessary to directly measure the strain of space in order to
detect gravitational radiation, just as it is not necessary to directly
measure the electric field of a light wave, which may also be exceedingly
small, in order to be able to detect this wave. Instead, one can
directly measure the
\textit{power}
conveyed by a beam of light by means of bolometry, for
example. Likewise, if one were to succeed in completely back-converting this
milliwatt of GR wave power with high efficiency back into a milliwatt of EM
power at the receiver, this amount of
\textit{power}
would be easily detectable by
standard microwave techniques.


\section{Signal-to-noise considerations}

The signal-to-noise ratio expected for the Hertz-like experiment depends on
the current status of microwave source and receiver technologies. Based on
the experience gained from the experiment done on YBCO using existing
off-the-shelf microwave components \cite{Chiao2004-1,Chiao2004-2,Chiao2004-3},
we expect that we would
need geometric-sized cross sections and a minimum conversion efficiency on
the order of
parts per million per transducer
to detect a signal. The overall system's signal-to-noise ratio depends on the
initial microwave power, the scattering cross section, the conversion
efficiency of the quantum transducers, and the noise temperature of the
microwave receiver (i.e., its first-stage amplifier).

Microwave low-noise amplifiers can possess noise temperatures that are
comparable to room temperature (or even better, such as in the case of
liquid-helium-cooled paramps or masers used in radio astronomy). The minimum
power $P_{\min }$ detectable in an integration time $\tau $ is given by%
\begin{equation}
P_{\min }=\frac{k_{\text{B}}T_{\text{noise}}\Delta \nu }{\sqrt{\tau \Delta \nu }}
\end{equation}%
where $k_{\text{B}}$ is Boltzmann's constant, $T_{\text{noise}}$ is the noise
temperature of the first stage microwave amplifier, and $\Delta \nu $ is its
bandwidth. Assuming an integration time of 1 sec, a bandwidth of 1
GHz, and a noise temperature of $T_{\text{noise}}=300$ K, one gets $P_{\min
}(\tau =$1 sec$)=1.3\times 10^{-16}$ W, which is much less than the
milliwatt power levels of typical microwave sources.


\section{Possible applications}

If we should be successful in the Hertz-like experiment, this could lead to
important possible applications in science and engineering. In science, it
would open up the possibility of gravitational wave astronomy at microwave
frequencies. One important problem to explore would be observations of the
analog of the cosmic microwave background (CMB) in gravitational radiation.
Because the universe is much more transparent to gravitational waves than to
electromagnetic waves, such observations would allow a much more penetrating
look into the extremely early Big Bang toward the Planck scale of time
than the currently well-studied CMB. Different cosmological models of the
very early universe give widely differing predictions of the spectrum of
this penetrating radiation, so that by measurements of the spectrum, one
could tell which model, if any, is close to the truth \cite{NASA2006}. The
anisotropy in this radiation would also be very important to observe.

In engineering, it would open up the possibility of intercontinental
communication by means of microwave-frequency gravitational waves directly
through the interior of the Earth, which is transparent to such waves. This
would eliminate the need of communications satellites and would allow
communication with people deep underground or underwater in submarines in
the oceans. Wireless power transmission by gravitational microwaves would
also be a possibility. Such a new direction of gravitational wave engineering
could aptly be called
``gravity radio."


\section[Appendix A]{Appendix A: \\
Wald's derivation of the Maxwell-like equations}

The Maxwell-like equations (Eqs. (\ref {Maxwell-like-eq-1})--(\ref{Maxwell-like-eq-4}))
are a consequence of the derivation by Wald
(see Ref. \cite{Wald})
in section 4.4, which starts from the assumption that for
weak gravitational fields, the metric of spacetime can be approximated
by (in the notation of
Misner, Thorne, and Wheeler
\cite{MTW})%
\begin{equation}
g_{\mu \nu }\approx \eta _{\mu \nu }+h_{\mu \nu }
\end{equation}%
where $g_{\mu \nu }$ is the metric tensor, $\eta _{\mu \nu }$ is the
Minkowski metric tensor for a flat spacetime, and $h_{\mu \nu }$ are small
perturbations of the metric tensor, such as those arising from gravitational
radiation.

When the lowest order effects of the motion of the source are
taken into account, but neglecting stresses, the linearized Einstein field
equations, when also linearized in the
nonrelativistic
velocity of the matter, become (in
units where $G=c=1$)%
\begin{equation}
\partial ^{\mu }\partial _{\mu }\overline{h}_{0\lambda }=16\pi J_{\lambda }
\label{wave-eq-for-h_0i}
\end{equation}%
where $\overline{h}_{\mu \nu }=h_{\mu \nu }-\frac{1}{2}\eta _{\mu \nu }h$
and where $J_{\lambda }$ is the mass current density four-vector of the
source. If, following Wald, one defines the
``vector potential"
as:%
\begin{equation}
A_{\mu }\equiv -\frac{1}{4}\overline{h}_{\mu \nu }t^{\nu }
\end{equation}%
where $t^{\nu }$ is the four-velocity of a test particle (which for a
nonrelativistic particle is time-like), one obtains%
\begin{equation}
\partial ^{\mu }\partial _{\mu }A_{\lambda }=-4\pi J_{\lambda }
\end{equation}%
These equations are equivalent to the Maxwell-like equations and have the
form of Maxwell's equations in the Lorentz gauge, with the consequence that
the perturbations $\overline{h}_{0\lambda }$ propagate with precisely the
speed of light $c$, and not at the speed $c/2$.

In contrast to this, using
the PPN formalism,
Braginsky, Caves, and Thorne \cite{BCT-1977}
derived a set of Maxwell-like equations
that yielded a speed of $c/2$, and not the speed of light $c$, for
time-varying perturbations of the fields. This difference in speeds arises
from the fact that the PPN formalism describes the near fields as seen by an
observer close to the source, but Wald's formalism describes the far fields
as seen by an observer in an asymptotically flat spacetime far away from the
source.

The standard transverse-traceless (TT) coordinate system can be
transformed into Wald's coordinate system by means of a local Galilean
coordinate transfomation \cite{Prague paper-1}. In the TT gauge, one of the
gauge conditions is%
\begin{equation}
h_{0\mu }\equiv 0
\end{equation}%
An incorrect conclusion drawn from this gauge condition is that only the
gravitoelectric components given by the strains $h_{ij}$ of a gravitational
plane wave exist, and that no gravitomagnetic components of radiation
fields in the far field of sources exist.
In \textit{Relativity on Curved Manifolds} (Chapter 9), Felice and Clarke
point out that the Riemann curvature
tensor for gravitational waves propagating in a flat background can be
separated into
``electric"
and
``magnetic"
parts, and that these Riemann
curvature tensor components satisfy tensor Maxwell-like equations \cite{FC-1973}.
The wave speed that follows from these equations is again precisely $c$. This
gauge-invariant way of characterizing gravitational radiation shows that the
``electric"
and
``magnetic"
components of the Riemann curvature tensor for a
monochromatic gravitational plane wave propagating in the vacuum are equal
in magnitude to each other in natural units.

The earliest mention of
Maxwell-like equations for linearized general relativity was perhaps made by
Forward in 1961 \cite{Forward}.


\section[Appendix B]{Appendix B: \\
Specular reflection of a circularly polarized EM wave
by a delocalized electron moving on a plane in the presence of a strong
magnetic field}
%

Here we address the question, what is the critical frequency for specular
reflection of an EM plane wave normally incident on a plane, in which
electrons are moving in the presence of a strong B field?
The motivation
for solving this problem is to answer also the following questions: How can
just a single electron on the outside surface of a
``Millikan oil drop"
generate enough current in response to
an incident EM wave, so as to produce a reradiated wave that totally
cancels out the incident wave within the interior of the drop, with the
result that none of the incident radiation can enter into the drop? Why does
specular reflection occur from the surface of such a drop, and hence why
does a hard-sphere EM cross section result for a pair of
``Millikan oil drops"?

To simplify this problem to its bare essentials, let us examine first a
simpler, planar problem consisting of a uniform electron gas moving
classically on a frictionless, planar dielectric surface. I start
from a three-dimensional point of view, but the Coulombic attraction of the electrons to
their image charges inside the dielectric will confine them in the direction
normal to the plane, so that the electrons are restricted to a two-dimensional
motion---that is, to frictionless motion in the two transverse dimensions of the plane.
The electrons are subjected to a strong DC magnetic field applied normally
to this plane. What is the linear response of this electron gas to a weak,
normally incident EM plane wave? Does a specular plasma-like reflection
occur below a critical frequency, even when  only a single, delocalized
electron is present on the plane? Let us first solve this problem
classically.

Let the plane in question be the $z=0$ plane, and let a strong, applied DC $%
\mathbf{B}$ field be directed along the positive $z$-axis. The Lorentz force
on an electron is given by%
\begin{equation}
\mathbf{F}=e\left( \mathbf{E}+\frac{\mathbf{v}}{c}\mathbf{\times B}\right)
\end{equation}%
where $\mathbf{E}$, the weak electric field of the normally incident plane
wave, lies in the $(x,y)$ plane. (I use Gaussian units only here in
this appendix.) The cross product $\mathbf{v\times B}$ is given by%
\begin{equation}
\mathbf{v\times B=}\left\vert
\begin{array}{ccc}
\mathbf{i} & \mathbf{j} & \mathbf{k} \\
v_{x} & v_{y} & 0 \\
0 & 0 & B%
\end{array}%
\right\vert =\mathbf{i}v_{y}B-\mathbf{j}v_{x}B
\end{equation}%
Hence, Newton's equations of motion reduce to $x$ and $y$ components only:%
\begin{equation}
F_{x}=m\ddot{x}=eE_{x}+\frac{v_{y}}{c}eB=eE_{x}+\frac{\dot{y}}{c}eB
\label{Newton-x}
\end{equation}%
\begin{equation}
F_{y}=m\ddot{y}=eE_{y}-\frac{v_{x}}{c}eB=eE_{y}-\frac{\dot{x}}{c}eB
\label{Newton-y}
\end{equation}%
Let us assume that the driving plane wave is a weak monochromatic wave with
the exponential time dependence%
\begin{equation}
E=E_{0}\exp \left( -i\omega t\right)
\end{equation}%
Then assuming a linear response of the system to the weak incident EM wave,
the displacement, velocity, and acceleration of the electron all have the
same exponential time dependence%
\begin{equation}
x=x_{0}\exp \left( -i\omega t\right) \text{ and }y=y_{0}\exp \left( -i\omega
t\right)
\end{equation}%
\begin{equation}
\dot{x}=\left( -i\omega \right) x\text{ and }\dot{y}=\left( -i\omega \right)
y
\end{equation}%
\begin{equation}
\ddot{x}=-\omega ^{2}x\text{ and }\ddot{y}=-\omega ^{2}y
\end{equation}%
which converts the two ODEs, Equations (\ref{Newton-x}) and (\ref{Newton-y}%
), into the two algebraic equations for $x$ and $y$%
\begin{equation}
-m\omega ^{2}x=eE_{x}-\frac{i\omega y}{c}eB
\end{equation}%
\begin{equation}
-m\omega ^{2}y=eE_{y}+\frac{i\omega x}{c}eB
\end{equation}%
Let us now add $\pm i$ times the second equation to the first equation.
Solving for $x\pm iy$, one gets%
\begin{equation}
x\pm iy=e\left( \frac{E_{x}\pm iE_{y}}{-m\omega ^{2}\pm \omega eB/c}\right)
\end{equation}%
where the upper sign corresponds to an incident clockwise circularly
polarized EM and the lower sign to an anticlockwise one. Let us define as
a shorthand notation%
\begin{equation}
z_{\pm }\equiv x\pm iy
\end{equation}%
as the complex representation of the displacement of the electron. Solving
for $z_{\pm }$, one obtains%
\begin{equation}
z_{\pm }=\frac{eE_{\pm }}{-m\left( \omega ^{2}\mp \omega \omega _{\text{cycl}%
}\right) }
\end{equation}%
where the cyclotron frequency $\omega _{\text{cycl}}$ is defined as%
\begin{equation}
\omega _{\text{cycl}}\equiv \frac{eB}{mc}
\end{equation}%
and where%
\begin{equation}
E_{\pm }\equiv E_{x}\pm iE_{y}
\end{equation}%
For a gas of electrons with a uniform number density $n_{e}$, the
polarization of this medium induced by the weak incident EM wave is given by%
\begin{equation}
P_{\pm }=n_{e}e\left( x\pm iy\right) =n_{e}ez_{\pm }=\frac{n_{e}e^{2}E_{\pm }%
}{-m\left( \omega ^{2}\mp \omega \omega _{\text{cycl}}\right) }=\chi
_{e}E_{\pm }
\end{equation}%
where the susceptibility of the electron gas is given by%
\begin{equation}
\chi _{e}=\frac{n_{e}e^{2}}{-m\left( \omega ^{2}\mp \omega \omega _{\text{%
cycl}}\right) }=-\frac{\omega _{\text{plas}}^{2}/4\pi }{\omega ^{2}\mp
\omega \omega _{\text{cycl}}}
\end{equation}%
where the plasma frequency $\omega _{\text{plas}}$ is defined by%
\begin{equation}
\omega _{\text{plas}}\equiv \sqrt{\frac{4\pi n_{e}e^{2}}{m}}
\end{equation}%
The index of refraction of the gas $n(\omega )$ is given by%
\begin{equation}
n(\omega )=\sqrt{1+4\pi \chi _{e}(\omega )}=\sqrt{1-\frac{\omega _{\text{plas%
}}^{2}}{\omega ^{2}\mp \omega \omega _{\text{cycl}}}}
\end{equation}%
Specular reflection occurs when the index of refraction becomes a pure
imaginary number. Let us define the critical frequency $\omega _{\text{%
crit}}$ as the frequency at which the index vanishes, which occurs when%
\begin{equation}
\frac{\omega _{\text{plas}}^{2}}{\omega _{\text{crit}}^{2}\mp \omega _{\text{%
crit}}\omega _{\text{cycl}}}=1
\end{equation}%
Because the index vanishes at this critical frequency, the Fresnel reflection
coefficient ${\mathcal{R}}(\omega )$ from the planar structure for normal
incidence at the critical frequency is given by%
\begin{equation}
{\mathcal{R}}(\omega )=\left\vert \frac{n(\omega )-1}{n(\omega )+1}%
\right\vert ^{2}\rightarrow 100\%\text{ when }\omega \rightarrow \omega _{%
\text{crit}}  \label{Fresnel-Reflection-for-plasma}
\end{equation}%
which implies specular reflection of the incident plane EM wave from the
electron gas. This yields a quadratic equation for $\omega _{\text{crit}}$,%
\begin{equation}
\omega _{\text{crit}}^{2}\mp \omega _{\text{crit}}\omega _{\text{cycl}%
}-\omega _{\text{plas}}^{2}=0
\end{equation}%
The solution for $\omega _{\text{crit}}$ is%
\begin{equation}
\omega _{\text{crit}}=\frac{\pm \omega _{\text{cycl}}\pm \sqrt{\omega _{%
\text{cycl}}^{2}+4\omega _{\text{plas}}^{2}}}{2}  \label{quadratic-solution}
\end{equation}%
The first $\pm $ sign is physical and is determined by the sense of
circular polarization of the incident plane wave. The second $\pm $ sign is
mathematical and originates from the square root. One of the latter
mathematical signs is unphysical. To determine which choice of the latter
sign is physical and which is unphysical, let us first consider the limiting
case when the inequality%
\begin{equation}
\omega _{\text{cycl}} \ll \omega _{\text{plas}}
\end{equation}%
holds. This inequality corresponds physically to the situation when the
magnetic field is very weak but the electron density is very high, so that
the phenomenon of specular reflection of EM waves with frequencies below the
plasma frequency $\omega _{\text{plas}}$ occurs. Let us therefore take the
limit $\omega _{\text{cycl}}\rightarrow 0$ in the solution given by Equation (\ref%
{quadratic-solution}). Negative frequencies are unphysical, so that we must
choose the positive sign in front of the surd as the only possible physical
solution. Thus, in general, it must be the case that the physical root of the
quadratic is given by%
\begin{equation}
\omega _{\text{crit}}=\frac{\pm \omega _{\text{cycl}}+\sqrt{\omega _{\text{%
cycl}}^{2}+4\omega _{\text{plas}}^{2}}}{2}  \label{Physical-root}
\end{equation}%
\qquad

Let us now focus on the more interesting case in which the magnetic field is
very strong but the number density of electrons is very small, so that the
plasma frequency is very low, corresponding to the inequality
\begin{equation}
\omega _{\text{cycl}} \gg \omega _{\text{plas}}
\end{equation}%
There are then two possible solutions, corresponding to clockwise-polarized
and anticlockwise-polarized EM waves, respectively, viz.,%
\begin{equation}
\omega _{\text{crit,1}}=\omega _{\text{cycl}}\text{ and }\omega _{\text{%
crit,2}}=0  \label{2-physical-solutions}
\end{equation}%
Note the important fact that these solutions are independent of the number
density (or plasma frequency) of the electron gas, which implies that even a
very dilute electron gas system can give rise to specular reflection. The
fact that these solutions are independent of the number density also implies
that they would apply to the case of an inhomogeneous electron density, such
as that arising for a single delocalized electron confined to the vicinity
of the plane $z=0$ by the
Coulomb
attraction to its image. Both solutions
of the quadratic Equation (\ref{2-physical-solutions}) are now physical
ones and imply that whether the sense of rotation of the EM polarization
corotates or counterrotates with respect to the magnetic-field-induced
precession of the guiding center motion of the electron around the magnetic
field determines which sense of circular polarization is transmitted when $%
\omega >$ $\omega _{\text{crit,2}}=0$, or which sense of circular
polarization is totally reflected when $\omega <\omega _{\text{crit,1}%
}=\omega _{\text{cycl}}$, provided that the frequency of the incident
circularly polarized EM wave is less than the cyclotron frequency $\omega _{%
\text{cycl}}$. The interesting solution is the one with the nonvanishing
critical frequency, because it implies that one solution always exists
where there is specular reflection of the EM wave, even when the number
density of electrons is extremely low (i.e., even when the plasma frequency $%
\omega _{\text{plas}}$ approaches zero), and even when this number density
becomes very inhomogeneous as a function of $z$.

In the extreme case of a single electron completely delocalized on the
outside surface of superfluid helium, one should solve the problem quantum
mechanically, by going back to Landau's solution of the motion of an
electron in a uniform magnetic field and adding as a time-dependent
perturbation the weak (classical) incident circularly polarized plane wave.
However, the above classical solution should hold in the correspondence
principle limit, where, for the single delocalized electron, the effective
number density of the above classical solution is determined by the absolute
square of the electron wave function, viz.,%
\begin{equation}
n_{e}=\left\vert \psi _{e}\right\vert ^{2}\text{ and}
\end{equation}%
\begin{equation}
\int n_{e}dV=\int \left\vert \psi _{e}\right\vert ^{2}dV=1
\end{equation}%
Here we must take into account the fact that there is a finite confinement
distance $d_{e}\approx $ 80 \AA\ in the $z$-direction of the electron's
motion in the hydrogenic ground state caused by the Coulomb attraction of
the electron to its image charge induced in the dielectric, but the electron
is completely delocalized in the $x$- and $y$-directions on an arbitrarily
large plane (and hence over the large spherical surface of a large drop).
The effective plasma frequency of the single electron may be extremely
small; nevertheless, total reflection by this single, delocalized electron
still occurs, provided that the frequency of the incident circularly
polarized EM wave is below the cyclotron frequency. The fundamental reason
why even just a single delocalized electron in a strong magnetic field can
give rise to specular reflection is that the $\mathbf{v\times B}$ Lorentz
force
leads to a longitudinal quantum Hall
resistance that is strictly zero, which shorts out the incident circularly
polarized EM wave. Thus, one concludes that the hard-wall boundary conditions
used in the order-of-magnitude estimate given by Equation (\ref%
{Geometric-X-section}) of the scattering cross section of microwaves from
the drops are reasonable ones. This conclusion will be tested experimentally
(see Fig.~\ref{fig04-Rescaled-scattering-experiment-4}).

The $\mathbf{v\times B}$ Lorentz force leads to a ``gravito-quantum Hall effect," in
which an electron, when subjected to a gravitational field $\mathbf{g}$ in a
quantum Hall sample, moves with a velocity that is perpendicular to both $%
\mathbf{g}$ and $\mathbf{B}$ fields. For example, an electron in a
vertically oriented, planar quantum Hall sample subjected to the Earth's
gravity field will move with a velocity at right angles to both the Earth's
$\mathbf{g}$ field and a horizontal DC $\mathbf{B}$ field applied normally
to the sample. This then induces a Hall current that is directly
proportional to, and perpendicular to, the applied $\mathbf{g}$ field.
Local, time-varying gravitational fields $\mathbf{g}(t)$ arising from a
gravitational wave impinging on the sample will induce time-varying
transverse \textit{electrical} currents in the quantum Hall sample in the DC
magnetic field. Since each electron carries mass as well as charge with it
when it moves, this radiation will also induce transverse, time-varying
\textit{mass} currents in this sample. The above analysis can be
generalized to gravitational waves, once the quadrupolar pattern of these
waves is taken into account. For one sense of circular polarization, a 180$%
^{\circ }$ phase shift between the transmitted and incident radiation fields
leads to the destructive interference of the transmitted and incident
radiation fields, independent of whether these fields are EM or GR in
nature. The destructive interference of the transmitted wave with the
incident wave in the forward direction leads to reflection of the incident
wave in the backward direction. The longitudinal quantum Hall resistance in
both EM and GR sectors vanishes, so that circularly polarized EM and GR
radiation fields of one sense are both ``shorted out,"
leading to the specular reflection for both kinds of waves.


\section[Appendix C]{Appendix C: \\
Tinkham's analysis of reflection from thin superconducting films}

It may be objected that Equations (\ref{Reflection-from-vacuum-superconductor-interface}), (\ref{Fresnel-reflection}), and (\ref{Fresnel-Reflection-for-plasma})
are believed to apply only when
the thickness $d$ of a sample is large compared with the relevant
penetration depth $\ell _{P}$, whereas the opposite limit (appropriate for a
thin-film sample) is assumed here. (In the case of superconductors, the
penetration depth $\ell _{P}$ is the London penetration depth $\lambda _{L}$%
.)

Contrary to this common belief, for the case in which the film is thin
compared to the penetration depth but the penetration depth is much
less than the radiation wavelength---that is,
$d \ll \ell _{P}$, but $\ell_{P} \ll \lambda $,
where $\lambda $ is the free space wavelength---the
reflectivity is not of the order of $(d/\ell _{P})^{2}$, as one might
naively expect. Rather, it is much higher, and in fact approaches unity as $\lambda
$ becomes infinite. See Equation (3.128) of Tinkham's book \cite{Tinkham}
for the transmissivity ${\mathcal{T}}$ of superconducting thin films, which
reads as follows:%
\begin{equation}
{\mathcal{T}}=\left[ \left( 1+\frac{\sigma _{1}Z_{0}d}{n+1}\right)
^{2}+\left( \frac{\sigma _{2}Z_{0}d}{n+1}\right) ^{2}\right] ^{-1}
\label{Tinkham's-(2.128)}
\end{equation}%
where $\sigma =\sigma _{1}+i\sigma _{2}$ is the complex conductivity of the
thin film, $d$ is its thickness, $n$ is the index of refraction of its
substrate, and $Z_{0}=\sqrt{\mu _{0}/\varepsilon _{0}}=\mu _{0}c$ is the
characteristic impedance of free space for EM waves. Although this equation
was derived by Tinkham in the context of superconductivity, it applies to
all thin films with a complex conductivity $\sigma =\sigma _{1}+i\sigma _{2}$%
. (It can also be readily generalized to the case of a complex conductivity
\emph{tensor}, which is applicable to the quantum Hall fluid.)

From this
equation, we see that the transmissivity can vanish in the low-frequency
limit $\omega \rightarrow 0$, since for superconductors $\sigma
_{2}\rightarrow 1/\omega \rightarrow \infty $, leading to a substantial
reflection of these waves
when there is
a negligible dissipation within the
superconducting film. This result can be understood in terms of an
inductance per square element of the thin film%
\begin{equation}
L=\mu _{0}\ell _{\text{gap}}
\end{equation}%
where $\ell _{\text{gap}}$ is a characteristic
energy-gap
length scale of the
superconductor or of the quantum Hall fluid. This leads to a reactance per
square element of the film of%
\begin{equation}
X_{L}=\omega L=\frac{1}{\sigma _{2}d}
\end{equation}%
whose low value is responsible for the high reflectivity for waves with
frequencies well below the relevant gap frequency. For details, see Refs.
\cite{Prague paper-1,Prague paper-2}.

However, in the derivation of Equation (3.128) in Tinkham's
book, it was assumed that the thin conducting film sample was transversely
infinite, so that it is not immediately obvious that it can be applied to
the electrons on a spherical
``Millikan oil drop,"
nor is it clear that the concept of a
``penetration depth"
applies to the
quantum Hall fluid on the surface of superfluid helium. Nevertheless, the
only relevant length scales for this fluid are the magnetic length scale (in
SI units) $\ell _{B}=(h/eB)^{1/2}$ for the quantum Hall effect and the
confinement distance scale $d_{e}$ of electrons on the superfluid drop
surface discussed in
Appendix B,
both of which are on the order of 10 nm
\cite{Prange,Grimes2-1,Grimes2-2}
(also see the footnote on page \pageref{Grimes2-fn}),
whereas the radius of a typical drop is around
4 mm, which is much larger than both of these microscopic length scales.

Because a small patch on the surface of a large spherical drop looks planar on
these length scales, one can still apply locally to this small patch, in the
limit of long wavelengths $\lambda $, the discontinuous-jump boundary
conditions for the tangential magnetic field that follows from the Maxwell
equation $\mathbf{\nabla \times B}=\mu _{0}\mathbf{J}$ and from its
gravitational analog $\mathbf{\nabla \times B}_{G}=\mu _{G}\mathbf{J}_{G}$.
It is these discontinuous-jump boundary conditions for the tangential
components of both $\mathbf{B}$ and $\mathbf{B}_{G}$ that lead to
nonnegligible reflections of both EM and GR waves from the quantum Hall
fluid on the surface of a drop. They are also the basis for Equation (3.128)
in Tinkham's book.

Therefore, the planar model used in the derivation of
Equation (3.128) in Tinkham's book should be valid for the reflectivity of
the spherical
``Millikan oil drops"
being
considered for the proposed experiment.
See Appendix B
for a discussion of the physical origin of the surface
currents responsible for the reflection in the case of GR waves. In the case
of EM waves, the transmissivity of EM waves at low frequencies is given by%
\begin{equation}
{\mathcal{T}}\approx 4\left( \frac{\omega L}{Z_{0}}\right) ^{2}=4\left(
\frac{\omega \mu _{0}\ell _{\text{gap}}}{\mu _{0}c}\right) ^{2}=4\left(
\frac{2\pi \ell _{\text{gap}}}{\lambda }\right) ^{2}
\label{Transmissivity-in-low-frequency-limit}
\end{equation}%
where the approximation has been made that $n\approx 1$. Thus, ${\mathcal{T}}$ %
is on the order of $(\ell _{\text{gap}}/\lambda )^{2}=(\omega /\omega _{%
\text{gap}})^{2}\approx (\omega /\omega _{\text{cycl}})^{2}$, since $\omega
_{\text{gap}}\approx \omega _{\text{cycl}}$ in the case of the quantum Hall
fluid. (See
Appendix B.)
Thus, the transmission ${\mathcal{T}}$ both of a
superconducting thin film and of a quantum Hall fluid film remains small,
and therefore the reflectivity $\mathcal{R}=1-{\mathcal{T}}$ of these films
remains high for all frequencies $\omega $ of an incident wave that are
well below the relevant gap frequency $\omega _{\text{gap}}$.

Note that the
permeability of free space $\mu _{0}$ cancels out of Equation (\ref%
{Transmissivity-in-low-frequency-limit}) and therefore that $\mu _{G}$ will
also cancel out of the analogous expression for the case of GR waves.
Therefore, since the quantum Hall fluid is strictly dissipationless,
a nonnegligible reflectivity results for both EM and GR waves from the
``Millikan oil drops"
for waves with
frequencies well below the relevant gap frequency---that is, the cyclotron
frequency $\omega _{\text{cycl}}$.

I thank an anonymous referee for pointing
out to me Equation (3.128) of Tinkham's book \cite{Tinkham}.


\section[Appendix D]{Appendix D: \\
How can a spin-1 photon be converted into a spin-2 graviton?}


The question of how a graviton (spin-2) can be produced from a photon (spin-1)
is important to consider. (I must thank Tom Kibble for raising this
important
question.)

The principle of equivalence should apply to all
charges and fields in curved spacetime \cite{Landau,Lamb-medal}.
However, Maxwell's equations for standard electromagnetism are expressed in
terms of fields on a \emph{flat} spacetime. They must be generalized to
fields on a \emph{curved} spacetime when interactions with gravitational
radiation are considered.

The back-action of EM waves propagating in a
curved spacetime on GR waves can in principle arise from the contribution
of the Maxwell stress-energy tensor, which is \emph{quadratic} in the EM
field strengths, as a source term on the right-hand side of Einstein's field
equations. In the absence of DC fields, such quadratic terms would give rise
to second harmonic generation in the conversion of EM to GR waves, but not
to first harmonic generation. However, there can in principle arise a \emph{%
linear} coupling of EM to GR waves when a DC magnetic (or DC electric) field
is present, and Einstein's equations are linearized in the weak EM and GR
wave amplitudes. This linear coupling can arise from a cross term, which
consists of a product of the DC field strength and the EM wave amplitude in
the quadratic Maxwell stress-energy tensor that leads to first harmonic
generation of GR waves at the same frequency as that of the incident EM
waves in a linear scattering process.

The role of the
``Millikan oil drops"
is that they can greatly enhance the
coupling between EM and GR waves due to their hard-wall boundary conditions
and mesoscopic gravitational masses. The electrons on their surfaces tightly
tie the local \textbf{B} field lines to these drops, so that these lines are
firmly anchored to the drops. At very low temperatures when the system
remains adiabatically in the ground state, the \textbf{B} field lines and
the drops co-move rigidly together according to a distant observer when the
system is disturbed by the passage of a GR or an EM wave. A given drop,
however, remains at rest with respect to a local inertial observer at the
center of the drop, and the local \textbf{B} field lines also do not appear
to move with respect to this local inertial observer. By contrast, to the
distant inertial observer in an asymptotically flat region of spacetime far
away from the pair of drops, where radiation fields become asymptotically
well defined, the two objects appear to be in relative motion, and the
system emits power in both GR and EM radiations.

Thus, a graviton (spin-2)
can in principle be produced from a photon (spin-1) in the presence of a DC
magnetic or a DC electric field (spin-1), in a scattering process from the
two objects. See
Ref. \cite{Halpern-1967} for
a quantum field theoretic treatment of such scattering processes.


%
\section*{Acknowledgments}
%
I thank John Barrow, Fran\c{c}ois
Blanchette, George Ellis, Sai Ghosh, Dave Kelley, Tom Kibble, Steve Minter,
Kevin Mitchell, James Overduin, Richard Packard, Jay Sharping, Martin
Tajmar, Kirk Wegter-McNelly, Roland Winston, and Peter Yu for
helpful discussions.


%
\end{document}